# Crystal Structure and Magnetic Properties of the $Ba_3TeCo_3P_2O_{14}$, $Pb_3TeCo_3P_2O_{14}$, and $Pb_3TeCo_3V_2O_{14}$ Langasites

J. W. Krizan[1*], C. de la Cruz[2], N. H. Andersen[3], and R. J. Cava[1]

[1]Department of Chemistry, Princeton University, Princeton, NJ 08544, USA

[2]Quantum Condensed Matter Division, Oak Ridge National Laboratory, Oak Ridge, Tennessee 37831-6393, USA

[3]Department of Physics, Technical University of Denmark, 2800 Kgs. Lyngby, Denmark

* Corresponding Author: jkrizan@princeton.edu

**Abstract**

We report the structural and magnetic characterizations of $Ba_3TeCo_3P_2O_{14}$, $Pb_3TeCo_3P_2O_{14}$, and $Pb_3TeCo_3V_2O_{14}$, compounds that are based on the mineral Dugganite, which is isostructural to Langasites. The magnetic part of the structure consists of layers of $Co^{2+}$ triangles. Nuclear and magnetic structures were determined through a co-refinement of synchrotron and neutron powder diffraction data. In contrast to the undistorted P321 Langasite structure of $Ba_3TeCo_3P_2O_{14}$, a complex structural distortion yielding a large supercell is found for both $Pb_3TeCo_3P_2O_{14}$ and $Pb_3TeCo_3V_2O_{14}$. Comparison of the three compounds studied along with the zinc analog $Pb_3TeZn_3P_2O_{14}$, also characterized here, suggests that the distortion is driven by $Pb^{2+}$ lone pairs; as such, the Pb compounds crystallize in a pyroelectric space group, P2. Magnetic susceptibility, magnetization, and heat capacity measurements were performed to characterize the magnetic behavior. All three compounds become antiferromagnetic with Néel temperatures $T_N \approx 21$ K ($Ba_3TeCo_3P_2O_{14}$), $\approx 13$ K ($Pb_3TeCo_3P_2O_{14}$), and $\approx 8$ K ($Pb_3TeCo_3V_2O_{14}$), and they exhibit magnetic transitions at high applied magnetic fields, suggesting intrinsically complex magnetic behavior for tetrahedrally coordinated $d^7$ $Co^{2+}$ in this structure type.

**Introduction**

Recently, Langasites have been the focus of extensive research due to their magnetic frustration,[1] multiferroic,[2] non-linear optical,[3] piezoelectric,[4] ferroelectric and dielectric properties.[4,5] Many of these phenomena result from the non-centrosymmetric space group (P321) in which they crystallize. Ferroelectricity is not directly possible in the P321 space group, but through the symmetry breaking by magnetic ordering or a structural phase transformation to one of the polar sub-groups, this property can be realized.[6] Some Langasites are available commercially for use in surface acoustic wave (SAW) filters due to their exceptional piezoelectric properties. Hundreds of compounds with the Langasite crystal structure have been reported. The four crystallographic positions occupied by the cations are significantly different and tolerant to the accommodation of a wide variety of metals, which makes the Langasites ideal for structure-property studies. Dugganite, $Pb_3TeZn_3As_2O_{14}$, is a mineral isostructural to Langasite, and is the inspiration for the compounds studied here.[7,8]

The prototypical Langasite, or $Ca_3Ga_2Ge_4O_{14}$ structure type (typified by $Ba_3TeCo_3P_2O_{14}$, Figure 1), possesses two-dimensional lattices of ions which are symmetry confined by multiple independent threefold rotation axes to be in equilateral triangles. Occupied by the appropriate ions,[9] these materials are ideal for the study of magnetic frustration due to the high symmetry of the magnetic interactions. The structure consists of two alternating layers of cations with triangle-based lattices: the first is composed of $AO_{10}$ decahedra and $MO_6$ octahedra, and the second of larger $M'O_4$ and smaller $M''O_4$ tetrahedra. The large tetrahedral site can be populated by a magnetic ion, forming a planar triangular cluster of $M'O_4$ tetrahedra sharing corners with the $MO_6$ octahedra in adjacent layers and the $M''O_4$ tetrahedra within the layer. This results in a magnetic lattice of two different size triangles. Alternatively, a magnetic rare earth can be placed in the decahedra to form a slightly puckered kagomé lattice. Interest in the magnetism of the Langasites has been primarily focused on materials with praseodymium or neodymium on the kagomé net, or others with iron in the triangular clusters (in each case with the other ions non-magnetic).[1,2,5,10–14] With interplay between magnetic frustration and multiferroicity, materials with cobalt in this structural motif are also currently of interest.[15]

Here we report our study of three compounds, $Ba_3TeCo_3P_2O_{14}$, $Pb_3TeCo_3P_2O_{14}$, and $Pb_3TeCo_3V_2O_{14}$, whose existence was described in a broad chemical survey of Langasites[7] - one of which having recently been characterized from a magnetic standpoint.[15] We report the results of structural determinations based on co-refinement of high resolution synchrotron powder X-ray diffraction and neutron powder diffraction data of both the nuclear and magnetic structures. The Ba compound is essentially a normal Langasite, but structural distortions are found to give rise to a large supercell in the Pb compounds, placing them in a pyroelectric space group (indicating that multiferroicity would be allowed by symmetry in these materials).[6] Characterization includes the field and temperature dependent magnetization and low temperature specific heat, which all suggest complex magnetic behavior.

**Experimental**

Polycrystalline samples were prepared by typical solid state methods, with slight modifications for each specific compound. Starting materials used for $Pb_3TeCo_3V_2O_{14}$ were PbO (Aldrich 99.999%), $TeO_2$ (99.995%), $Co_3O_4$ (99.7%), and $V_2O_5$ (Alfa >99.6%). Starting materials for $Pb_3TeCo_3P_2O_{14}$ and $Ba_3TeCo_3P_2O_{14}$ were PbO, $BaCO_3$, $TeO_2$, $Co_3O_4$, and $NH_4H_2PO_4$ (Fisher, HPLC Grade). The non-magnetic analogues $Pb_3TeZn_3P_2O_{14}$, $Pb_3TeZn_3V_2O_{14}$ and $Ba_3TeZn_3P_2O_{14}$ were synthesized from PbO, $BaCO_3$, $TeO_2$, ZnO (Alfa 99.99%), $V_2O_5$, and $NH_4H_2PO_4$ for use as background subtraction materials in the specific heat studies. Stoichiometric (by metal) mixtures were ground and placed in high density, covered alumina crucibles and pre-reacted overnight at 550 °C ($Pb_3TeCo_3V_2O_{14}$) or 450 °C under flowing oxygen. $Pb_3TeCo_3V_2O_{14}$ was fully reacted after subsequent heating at 625 °C in air. $Pb_3TeCo_3P_2O_{14}$ was synthesized by additional heating for 146 hours at 700 °C in a covered crucible under flowing oxygen with intermittent grindings. $Ba_3TeCo_3P_2O_{14}$ was synthesized by subsequent heating at 800 °C for 120 hours. The zinc analogues were synthesized under identical conditions except for $Ba_3TeZn_3P_2O_{14}$, which was heated at 900 °C; all were heated until shown to be pure by laboratory X-ray powder diffraction (XRD). Laboratory diffraction experiments were carried out on Rigaku Miniflex II, and Bruker D8 Focus diffractometers, both Cu Kα with graphite monochromator. Low temperatures and an oxidizing atmosphere were critical in minimizing the volatilization of the starting materials and the formation of a pure product. $Pb_3TeCo_3P_2O_{14}$ and $Ba_3TeCo_3P_2O_{14}$ powders are very bright indigo and purple respectively. $Pb_3TeCo_3V_2O_{14}$ is a dull slate color with a hint of blue.

Bulk magnetic measurements were performed on powders of mass 75 – 140 mg using a CRYOGENIC Cryogen Free Measurement System (CFMS) Vibrating Sample Magnetometer (VSM). The magnetic susceptibility was recorded in a field of 0.1 T and temperatures between 2 and 300 K, using a scanning mode with a ramping rate of 2 K/min. Magnetization measurements were performed at three temperatures: 2 K, 11 K, and 23 K ($Ba_3TeCo_3P_2O_{14}$) or 30 K ($Pb_3TeCo_3P_2O_{14}$ and $Pb_3TeCo_3V_2O_{14}$), and fields up to 16 T using scanning sequences with a ramping rate of 0.2 T/min. Heat capacity measurements between 2 K and 250 K were conducted in a Quantum Design Physical Property Measurement System using the heat relaxation method. Dense polycrystalline pellets of 5-10 mg were placed on the sapphire platform using Apiezon N grease, of which the heat capacity was previously measured for subtraction. Synchrotron X-ray powder diffraction (SXRD) data was collected at the Advanced Photon Source at Argonne National Lab on beamline 11-BM. The Pb-containing samples were loaded into 0.1 mm diameter glass capillaries to reduce sample absorption. $Ba_3TeCo_3P_2O_{14}$ was loaded in a 0.8 mm Kapton tube due to its lower predicted absorption. The absorption of the materials was calculated using the Cromer & Liberman algorithm through the APS web interface.[16] SXRD patterns were collected at a wavelength of 0.413159 Å over a 2θ range of 0.5 to 50 degrees at 100 K to minimize the thermal parameters and match the temperature of the neutron diffraction data.

Neutron powder diffraction (NPD) patterns for $Ba_3TeCo_3P_2O_{14}$ and $Pb_3TeCo_3P_2O_{14}$ were collected at the High Flux Isotope Reactor beam line HB-2A at Oak Ridge National Laboratory. Diffraction patterns were collected at both 100 K and 4 K with 2.41 Å and 1.54 Å wavelength neutrons, as produced by Ge[113] and Ge[115] reflections from a vertically focusing germanium wafer-stack monochromator. This resulted in four NPD data sets for each compound. For all experiments, the collimation was set to 12'-open-6' to increase the neutron flux but at the detriment of the resolution; this improved the statistics for the determination of the magnetic structure and increased the visibility of the superstructure reflections. The magnetic structures were determined and refined through the use of the SARAh and FullProf software suites.[17–19]

**$Ba_3TeCo_3P_2O_{14}$ Structural Analysis**

The diffraction patterns for $Ba_3TeCo_3P_2O_{14}$ showed an ideal Langasite structure, space group P321. Peak shapes were modeled with the Thompson-Cox-Hastings pseudo-Voight profile convoluted with axial divergence asymmetry, but the high resolution SXRD pattern revealed an additional asymmetry for many of the peaks (Figure 2). Using the prototypical Langasite structure, this resulted in a good fit to the data. However, to account for the additional peak shape asymmetry which may be due to anisotropic strain arising at the grain boundaries,[20] the SXRD pattern was also modeled in FullProf as a convolution of two identical phases with very slightly different cell parameters. There is no indication that two distinct phases are actually present; even at the highest angles in the high resolution SXRD pattern there is a continuous distribution of intensity. For the final refinements of the structure, five diffraction patterns were co-refined simultaneously: four neutron diffraction patterns (Figure 3; at 2.41 Å and 1.54 Å at both 100 and 4 K) were simultaneously refined along with the SXRD data taken at 100 K, using the same Debye-Waller factor for both temperatures. All atomic positions were refined. For the 4 K data, the magnetic structure was also refined; this is described in a later section. The fit of the model

Krizan 5

to the SXRD and NPD patterns is excellent (Figures 2 and 3). The results of this refinement are presented in Table 1.

**$Pb_3TeCo_3P_2O_{14}$ and $Pb_3TeCo_3V_2O_{14}$ Structural analysis**

Analysis of the SXRD patterns indicated that the basic Langasite structure with trigonal symmetry could account for many of the peaks in the Pb-based compounds,[6] but that there were many superlattice peaks present (Figure 4). Beginning with the nonisomorphic subgroups of space group P321, and the subsequent subgroups, it was determined that the space group of the supercell is P2 for both $Pb_3TeCo_3P_2O_{14}$ and $Pb_3TeCo_3V_2O_{14}$.

For $Pb_3TeCo_3P_2O_{14}$, iterative structure solutions using only the SXRD pattern were first performed by varying only the positions of the cations, first in the prototypical P321 cell, then in the larger, lower symmetry P2 cell. The very large unit cell and the number of independent atoms required that chemically reasonable constraints be imposed to facilitate the full superstructure refinement. Tellurium and phosphorous atoms were initially refined as the centers of octahedral $TeO_6$ and tetrahedral $PO_4$ rigid body groups (RBG), respectively. RBGs are hard constraints on the positions of the coordinating oxygen atoms in relation to the cations and to each other. These groups can then translate and rotate as a unit centered on the cation. In the octahedral $TeO_6$ RBG, the three oxygen atoms above the tellurium are symmetry related to those below. This was initially set with a perfect octahedral orientation with one face of the octahedron parallel to the *ab*-plane and a fixed Te-O bond length of 1.927 Å as determined from the initial refinements of $Ba_3TeCo_3P_2O_{14}$. The phosphorous-oxygen tetrahedron is well characterized in the literature and is known to be fairly rigid in the solid state.[21] The $PO_4$ tetrahedron was modeled after "orthophosphates proper" and adjusted for the apical oxygen (O*) corner sharing with the lead polyhedra, and for the base oxygens (O**) edge sharing with the lead polyhedra.[21] This resulted in an initial model for the phosphate tetrahedron with the apical P-O* (roughly parallel to the c-axis) distance of 1.53045 Å and a base P-O** distance of 1.546 Å. The bond angles between the apex and the base (O*-P-O**) and within the base (O**-P-O**) were set as 111.2° and 106.3°, respectively.

This model was implemented in a co-refinement of the supercell structure performed with the SXRD pattern at 100 K (0.413 Å) and the NPD patterns at 100 K and 4 K (both at 1.54 Å and 2.41 Å). Initial refinements allowed for the translation of the PO$_4$ tetrahedra in three dimensions, and for the three symmetry inequivalent oxygen atoms forming the octahedron around tellurium to rotate as a unit in the *ab*-plane. The fit was further improved by allowing for the tilting and twisting of the PO$_4$ tetrahedra. For the co-refinement of the five diffraction patterns (four neutron diffraction patterns, at 2.41 Å and 1.54 Å at both 100 and 4 K and the SXRD data taken at 100 K) atomic positions and thermal parameters were constrained to be the same for both the high temperature and low temperature neutron and SXRD data. To account for thermal contraction of the unit cell, the *a:b*-ratio and the monoclinic β angle were held constant in the neutron patterns since there was not sufficient resolution to distinguish the subtle differences in the temperature dependence of the metric distortion from trigonal symmetry. Careful analysis of the individual patterns indicated that no significant structural changes occurred between 5 and 100 K. Thus the superstructure reported can be considered a thermal average of the 5 K and 100 K structures; any differences in the thermal parameters that may be present are beyond the scope of this study.

To take full advantage of the better statistics of the 4 K neutron data, the magnetic contribution to the pattern needed to be modeled; alternating between refinement of the structural and magnetic components resulted in further improvements to the fit. The magnetic structure is described in a later section of this report. The initially-employed rigid body model for the P-O and Te-O polyhedra was disabled after achieving a reasonable fit to the magnetic structure and replaced by soft constraints on the M-O bond lengths and O-O distances (Table 2). The final structural model for Pb$_3$TeCo$_3$P$_2$O$_{14}$ is presented in Table 3. The excellence of the fit of the model to the data can be seen in the SXRD pattern shown in figure 4a, and in the NPD patterns shown in Figure 5.

In the case of $Pb_3TeCo_3V_2O_{14}$, neutron diffraction data was not collected due to previous work[15] and therefore only SXRD data was refined. The rigid body model developed for the full superstructure of $Pb_3TeCo_3P_2O_{14}$ was carried over to the vanadate case with a slight modification to account for the size difference between phosphorous-oxygen and vanadium-oxygen tetrahedra. The vanadium-oxygen bond lengths were changed to those typically observed,[22] but no changes in the orientation of the tetrahedra were allowed other than simple translation. The oxygen coordination around the tellurium atoms was not modified, but the position of the tellurium was refined. This yielded a stable refinement and an exceptionally good fit to the synchrotron diffraction pattern (Figure 4b) with few anomalous bond lengths. Metal oxide bond lengths, especially the Co-O bond lengths, are not expected to be well defined in this experiment.

The crystal structure of $Pb_3TeCo_3P_2O_{14}$ is compared to $Ba_3TeCo_3P_2O_{14}$ in Figures 6 and 7. Atomic positions of the Pb compounds are given in Table 3 and 4 respectively. Selected bond lengths relating to the Pb-O, Ba-O and Co-O coordination polyhedra in $Ba_3TeCo_3P_2O_{14}$ and $Pb_3TeCo_3P_2O_{14}$ are given in Table 5.

**$Pb_3TeZn_3P_2O_{14}$ Structure Analysis**

$Pb_3TeZn_3P_2O_{14}$, $Pb_3TeZn_3V_2O_{14}$, and $Ba_3TeZn_3P_2O_{14}$ were synthesized as non-magnetic analogs for subtracting the lattice contribution in heat capacity measurements and were characterized using laboratory PXRD. $Pb_3TeZn_3P_2O_{14}$ was found to be isostructural to $Pb_3TeCo_3P_2O_{14}$. Using the atomic coordinates obtained for the cobalt variant as a model results in an excellent fit to the laboratory XRD pattern for $Pb_3TeZn_3P_2O_{14}$ (see Figure 8). Only the cell dimensions and an overall Debye-Waller factor were refined. The monoclinic distortion puts the structure in space group P2 with unit cell parameters *a* = 14.51284(3), *b* = 25.1385(2), *c* = 5.1874(4), β = 90.0631 (β was fixed at the value found for Pb$_3$TeCo$_3$P$_2$O$_{14}$). The inset to Figure 8 shows details of the pattern in a region including superlattice reflections, which are visible in the laboratory powder XRD patterns, illustrating the quality of the fit. Final $\chi^2$ and $R$ values are presented for the final structural model in Table 6. For quantitative determination of the structure of this phase, datasets and refinements similar to those performed here for Pb$_3$TeCo$_3$P$_2$O$_{14}$ would be required.

**Results – Bulk Magnetic Properties**

Figure 9 shows the temperature dependent magnetic susceptibility and inverse susceptibility both on heating and cooling with the results of a Curie-Weiss fit, for all three compounds. The Curie Weiss fits are to the data on heating and are of the type $\chi = \frac{C}{T-\theta}$ where C is the Curie constant, and θ the Weiss temperature. All three samples order antiferromagnetically, with ordering temperatures of 6.5(1) K, 9.172(12) K, 12.9(1) K, and 20.9(4) K, for Pb$_3$TeCo$_3$V$_2$O$_{14}$ (two transitions), Pb$_3$TeCo$_3$P$_2$O$_{14}$, and Ba$_3$TeCo$_3$P$_2$O$_{14}$ respectively. These ordering temperatures were obtained from the maxima in a $\frac{d\chi T}{dT}$ vs. $T$ plot. For low applied fields these materials act as traditional antiferromagnets; the Co$^{2+}$ moments found in the fits to the susceptibility data above 150 K are 4.46, 4.73, and 4.69 μ$_B$, respectively. These are higher than the expected spin only value (3.87 μ$_B$) due to the presence of an incompletely quenched orbital contribution and agree with other experimental observations of Co$^{2+}$. The Weiss temperatures are -21.05 K, -30.60 K and -20.29 K, respectively.

The magnetic transitions were further characterized by the heat capacity, as shown in Figure 10. The magnetic contribution to the heat capacity was determined by subtracting the lattice component estimated by using non-magnetic Pb$_3$TeZn$_3$P$_2$O$_{14}$, Pb$_3$TeZn$_3$V$_2$O$_{14}$ and Ba$_3$TeZn$_3$P$_2$O$_{14}$ as standards. No scaling factor was employed, or necessary. Sharp ordering transitions were observed and all are in agreement with susceptibility data. Though Pb$_3$TeCo$_3$P$_2$O$_{14}$ and Ba$_3$TeCo$_3$P$_2$O$_{14}$ exhibit shoulders in the heat capacity at lower temperatures. Pb$_3$TeCo$_3$V$_2$O$_{14}$ exhibits two transitions in the heat capacity, in agreement with previously reported data,[15] indicating the presence of two magnetically ordered phases at zero applied field. In all cases, there is a significant amount of magnetic entropy released on cooling at temperatures above the 3D ordering transition. The total entropy released approaches the amount expected for spin 3/2, with the small differences from the expected value likely due to the inexactness of the lattice heat capacity subtraction.

Figure 11 shows the VSM measurements of the magnetizations at 3 K and 11 K for all three samples, at 30 K for $Pb_3TeCo_3P_2O_{14}$ and $Pb_3TeCo_3V_2O_{14}$, and at 23 K for $Ba_3TeCo_3P_2O_{14}$, recorded in increasing and decreasing fields up to 16 T. The results for $Pb_3TeCo_3P_2O_{14}$ and $Ba_3TeCo_3P_2O_{14}$ show pronounced hysteresis, and in particular $Ba_3TeCo_3P_2O_{14}$ displays a dramatic jump in the magnetizations at low temperatures, indicating first order field induced phase transitions. The effect of field on the magnetic states of $Ba_3TeCo_3P_2O_{14}$ is particularly dramatic; there are clearly three different magnetic ordering regimes encountered before the 16 T limit of the applied field. Although there is no significant hysteresis in the $Pb_3TeCo_3V_2O_{14}$ sample, it displays an S-shape variation at low temperatures that also indicates a magnetic transition driven by the applied field.

**Results – Magnetic Structures**

The magnetic superlattice reflections of $Ba_3TeCo_3P_2O_{14}$ were indexed with a k-vector of (1/3, 1/3, 1/2). This k-vector was determined through an iterative search through the Brillouin zone focusing on the high symmetry directions. The software package SARAh Refine was used for this purpose with a Reverse Monte Carlo algorithm to perform a rough fit to the data and determine the most likely k-vector. After successfully indexing the patterns, SARAh Representational Analysis was used to calculate the different symmetry allowed basis vectors (ψ) and the irreducible representations (Γ) that they could form.[19] Using the FullProf Software suite, the irreducible representations were subsequently analyzed to determine which could most accurately model the magnetic structure peaks before proceeding with a careful quantitative refinement.

The magnetic structures of $Ba_3TeCo_3P_2O_{14}$ and $Pb_3TeCo_3P_2O_{14}$, as expressed in terms of the irreducible representation (Γ) and basis vectors (ψ) in the scheme used by Kovalev's tabulated works,[23] are reported in Tables 7 and 8 respectively. For both $Pb_3TeCo_3P_2O_{14}$ and $Ba_3TeCo_3P_2O_{14}$ (Figure 12), the refinements of the magnetic structures at zero applied field indicate that there are ferromagnetic triangular cobalt clusters that are coupled antiferromagnetically both in the *ab*-plane and along the *c*-axis by nature of the respective k-vectors.

$Pb_3TeCo_3P_2O_{14}$, however, exhibits another type of cluster behavior. The cobalt atoms at $a=1/2$ must have moments along the *b*-axis as a result of the symmetry constraints allowed by $Γ_1$. A symmetry analysis places no restrictions on the direction of the adjacent (symmetry equivalent) moments in the triad and so to facilitate finding a simple model that can be used to approximate the true magnetic structure, these are assumed to be clusters with zero net spin. The model presented here in Figure 12 is the simplest model found that reasonably fits the data using $Γ_1$. While it is not immediately clear what would lead to this behavior, the modulation of the moments improves the fit and this behavior is consistent with the model presented here for $Ba_3TeCo_3P_2O_{14}$ and the model for $Pb_3TeCo_3V_2O_{14}$ in the literature.[15]

Both magnetic structures are consistent with the geometry of the oxygen sublattice. Looking at superexchange interactions within the triangular clusters, a nearest-neighbor ferromagnetic interaction is expected, and in the inter-triangle Co-O-O-Co case, super-superexchange interactions are mediated through a series of large bond angles that give the expected antiferromagnetic interaction. Interestingly however, both structures seem to illustrate mixed antiferromagnetic and ferromagnetic interactions between adjacent triangular clusters within the *ab*-plane, suggesting a competition between the two interactions.

**Discussion and Conclusion**

$Ba_3TeCo_3P_2O_{14}$ displays an essentially prototypical P321 Langasite structure. The phosphate tetrahedra display positional disorder in this compound, however, manifested in the split oxygen positions in the $PO_4$ units (Figure 1). In the refinement of $Pb_3TeCo_3P_2O_{14}$, the large cell posed a challenge for obtaining an accurate full superstructure determination from the powder data. With 78 unique atoms in the unit cell of $Pb_3TeCo_3P_2O_{14}$, spanning from very heavy to very light, co-refinement of neutron and X-ray data was needed for a successful determination of the full superstructure. Careful use of hard and soft constraints on the atomic positions was employed to reach a final structural model. After lifting the RBG restraints and replacing them with soft bond length constraints, both $TeO_6$ and $PO_4$ polyhedra retained their shapes and did not distort significantly. In applying the superstructure model derived for $Pb_3TeCo_3P_2O_{14}$ to $Pb_3TeCo_3V_2O_{14}$, it was hypothesized that the oxygen sublattices are almost but not quite identical. The stable refinement and the exceptionally good fit to SXRD data support this hypothesis. Co-refinement of SXRD and NPD data would improve the details of our structural model for $Pb_3TeCo_3V_2O_{14}$, rectifying some of the anomalous M-O bond lengths. We note that complex structural distortions of the $La_3SbZn_3Ge_2O_{14}$ Langasite have also been reported, with an even larger unit cell than that found here, but no structural solution was obtained.[24]

Comparing these three compounds shows the subtle effects of the stereochemical lone pairs on $Pb^{2+}$. The fact that both the $Pb^{2+}$ variants (and the Zn analogues) display a nearly identical metric distortion from trigonal symmetry suggests that the cation in the decahedral site has a strong influence on the rest of the structure. The Pb lone pairs are the driving force, and result in a transition metal independent effect on the crystal structure. This effect is seen in the comparison of $Pb_3TeCo_3P_2O_{14}$ and $Ba_3TeCo_3P_2O_{14}$ (Figure 6); it is clear that the introduction of Pb has resulted in significant atomic displacements. With respect to the Ba analog, the Pb analog displays its most significant deviations from the higher symmetry in the *ab*-plane. The most notable shifts are those of the Pb and the phosphate tetrahedra: the phosphate tetrahedra predominantly rotate in the *ab*-plane and the Pb cations immediately surrounding each tetrahedron rotate counter to this motion. The PO$_4$ tetrahedra share all oxygen anions with the surrounding Pb polyhedra resulting in an intimate relationship between the two. Distortions of the phosphate groups and the Pb coordination are thus likely both a result of the stereochemical lone pairs on Pb$^{2+}$.[25] Figure 7 compares the Pb-O coordination to the Ba-O coordination. Within their coordination spheres, the Pb and the Ba sit closer to the oxygens shared with the TeO$_6$ octahedra. However, this is significantly more exaggerated in the Pb-O coordination, in agreement with the expected involvement of the lone pairs. The range of bond lengths for the Pb-O coordination polyhedra are compared to those of the Ba-O polyhedra in Table 5.

Structurally, substituting Pb for Ba in Pb$_3$TeCo$_3$P$_2$O$_{14}$ and Pb$_3$TeCo$_3$V$_2$O$_{14}$ (and Pb$_3$TeZn$_3$P$_2$O$_{14}$) pushes these materials over a tipping point. The lone pairs on Pb$^{2+}$ push the structures into a large supercell that breaks the strict three-fold symmetry, with the consequence that some of the magnetic frustration intrinsic to the Langasite structure is relieved. With respect to the corresponding dimension in the original P321 cell, the *a*-axis in the monoclinic cell is compressed, resulting in symmetry-confined isosceles triangle cobalt clusters. Additionally, in both Pb$_3$TeCo$_3$P$_2$O$_{14}$ and Ba$_3$TeCo$_3$P$_2$O$_{14}$, the cobalt cations are displaced towards the edges of their tetrahedra. This can be considered as an overall contraction of the cobalt clusters, suggesting strong magnetic interactions between cobalt cations within each discrete cluster.

Dugganite and other structural analogues of Langasite with Pb$^{2+}$ in the decahedral site are likely to exhibit similar behavior, and should be further investigated by SXRD. (The previously reported structure of Pb$_3$TeCo$_3$V$_2$O$_{14}$[15] actually represents the substructure, determined without taking into account the superlattice.) In the compounds reported here, even taking the supercell into account in the current refinements, the characterization of the phosphorous tetrahedra indicates that in all cases some residual disorder is present; the refinements suggest that there is room for the small PO$_4$ tetrahedron to rattle in the cage formed by the rest of the structure and that the stereochemical lone pairs on the Pb$^{2+}$ help to direct the orientations of the tetrahedra. (The "disorder" of the tetrahedra could reflect the presence of a supercell that is even larger than the one detected and described here.)

The compounds all display typical Curie-Weiss behavior in low magnetic fields. In the case of $Pb_3TeCo_3V_2O_{14}$, this is in agreement with a previous report.[15] For each of the three materials studied, the magnetic transition is reasonably sharp. The index of frustration ($\theta/T_n$) is apparently low, but this is due to the low values of $\theta$ that artificially result from the competition between ferromagnetic and antiferromagnetic interactions in these compounds. Heat capacity data for all three compounds showed sharp transitions corresponding to the antiferromagnetic ordering. $Pb_3TeCo_3V_2O_{14}$ exhibited two sharp transitions, as previously reported. The magnetic heat capacities (Figure 10) show that both $Pb_3TeCo_3P_2O_{14}$ and $Ba_3TeCo_3P_2O_{14}$ have anomalies below the sharp ordering peak that could potentially be representative of broad transitions. This is further corroborated by slight kinks in the high resolution magnetic susceptibility (Figure 9) at the corresponding temperatures. These transitions could be related to the second sharp transition observed in $Pb_3TeCo_3V_2O_{14}$, though further work is needed to analyze this complex magnetic behavior. All three compounds released a significant amount of entropy at temperatures above the 3D magnetic ordering transition. This suggests that short range ordering is present above the 3D ordering temperatures. Of the three, the Ba analog is most interesting from this perspective – it releases half of the entropy before the 3D ordering transition on cooling, suggesting the presence of much more substantial short range order before the 3D $T_N$ than is exhibited by the other two compounds.

All three of the compounds display complex behavior in magnetic fields greater than 5 T. Up to this point, the magnetization is linear. At higher fields, several magnetic states are observed. While none of these compounds has a fully saturated magnetization by 16 T at 3 K, $Ba_3TeCo_3P_2O_{14}$ is nearly saturated. This compound shows at least two distinct magnetic transitions and a large hysteresis. $Pb_3TeCo_3P_2O_{14}$ and $Pb_3TeCo_3V_2O_{14}$ behave similarly, with one distinct transition occurring far before magnetization saturation; this suggests that there may be another transition in these compounds analogous to the one at high field observed in $Ba_3TeCo_3P_2O_{14}$, but just outside of the measurement range. These measurements also suggest that the magnetic behavior is similar in all the materials, a property of $Co^{2+}$ in this particular crystallographic environment and magnetic sublattice, but that the behavior may be different in the Ba analog due to its higher symmetry.

The structural distortions observed in $Pb_3TeCo_3P_2O_{14}$ and $Pb_3TeCo_3V_2O_{14}$ and their novel magnetic properties under applied field makes them of interest as potential multiferroics. Materials with cobalt in the $Ca_3Ga_2Ge_4O_{14}$ structure type appear to exhibit novel behavior under high magnetic fields, and thus it is of interest to look for more $Co^{2+}$ variants: few are reported compared to those with $Fe^{3+}$ on the same site. Further measurements of these materials will ideally involve single crystal measurements of both the crystal structure and the magnetism. The simple model of the magnetic structure reported here, while not unique and despite showing such a difference in the nature of the magnetism in the cobalt triads, exhibits some fundamental characteristics that are in agreement with the nuclear structure and the magnetic properties; single crystal experiments are needed to determine the magnetic structure unambiguously. To date, the magnetic Langasites and Dugganites have only been reported in the noncentrosymmetric, trigonal space group P321; the two Pb variants reported here are the first compounds with structures in one of the pyroelectric space groups, a necessary symmetry requirement for the rise of multiferroicity. Further measurements would be of interest to determine whether the materials studied here are indeed multiferroic.

**Acknowledgements**

The authors would like to thank S. Dutton for helpful discussions. This research was supported by the U. S. Department of Energy, Division of Basic Energy Sciences, Grant DE-FG02-08ER46544. The authors thank the 11-BM team for their excellent synchrotron diffraction data; use of the Advanced Photon Source at Argonne National Laboratory was supported by the U. S. Department of Energy, Office

**References**

(1) Choi, K. Y.; Wang, Z.; Ozarowski, A.; Van Tol, J.; Zhou, H. D.; Wiebe, C. R.; Skourski, Y.; Dalal, N. S. *Journal of Physics: Condensed Matter* **2012**, *24*, 246001.
(2) Zhou, H. D.; Lumata, L. L.; Kuhns, P. L.; Reyes, A. P.; Choi, E. S.; Dalal, N. S.; Lu, J.; Jo, Y. J.; Balicas, L.; Brooks, J. S.; Wiebe, C. R. *Chem. Mater.* **2009**, *21*, 156–159.
(3) Heimann, R. B.; Hengst, M.; Rossberg, M.; Bohm, J. *physica status solidi (a)* **2003**, *195*, 468–474.
(4) Mill, B. V.; Pisarevsky, Y. V. In *Frequency Control Symposium and Exhibition, 2000. Proceedings of the 2000 IEEE/EIA International*; IEEE, 2000; pp. 133–144.
(5) Marty, K.; Bordet, P.; Simonet, V.; Loire, M.; Ballou, R.; Darie, C.; Kljun, J.; Bonville, P.; Isnard, O.; Lejay, P.; Zawilski, B.; Simon, C. *Phys. Rev. B* **2010**, *81*, 054416.
(6) Hahn, T. *International Tables for Crystallography, Volume A: Space Group Symmetry*; 5th ed.; Wiley, 2011.
(7) Mill, B. V. *Russ. J. Inorg. Chem.* **2009**, *54*, 1205–1209.
(8) Anthony, J. W.; Bideaux, R. A.; Bladh, K. W. *Handbook of Mineralogy*; Nichols, M. C., Ed.; Mineralogical Society of America: Chantilly, VA 20151-1110, USA., 2006.
(9) Belokoneva, E. L.; Simonov, M. A.; Butashin, A. V.; Mill', B. V.; Belov, N. V. *Soviet Physics Doklady* **1980**, *25*, 954.
(10) Bordet, P.; Gelard, I.; Marty, K.; Ibanez, A.; Robert, J.; Simonet, V.; Canals, B.; Ballou, R.; Lejay, P. *Journal of Physics: Condensed Matter* **2006**, *18*, 5147.
(11) Ghosh, S.; Datta, S.; Zhou, H.; Hoch, M.; Wiebe, C.; Hill, S. *J. Appl. Phys.* **2011**, *109*, 07E137.
(12) Marty, K.; Simonet, V.; Bordet, P.; Ballou, R.; Lejay, P.; Isnard, O.; Ressouche, E.; Bourdarot, F.; Bonville, P. *Journal of Magnetism and Magnetic Materials* **2009**, *321*, 1778–1781.
(13) Loire, M.; Simonet, V.; Petit, S.; Marty, K.; Bordet, P.; Lejay, P.; Ollivier, J.; Enderle, M.; Steffens, P.; Ressouche, E.; Zorko, A.; Ballou, R. *Phys. Rev. Lett.* **2011**, *106*, 207201.
(14) Lumata, L. L.; Besara, T.; Kuhns, P. L.; Reyes, A. P.; Zhou, H. D.; Wiebe, C. R.; Balicas, L.; Jo, Y. J.; Brooks, J. S.; Takano, Y.; Case, M. J.; Qiu, Y.; Copley, J. R. D.; Gardner, J. S.; Choi, K. Y.; Dalal, N. S.; Hoch, M. J. R. *Phys. Rev. B* **2010**, *81*, 224416.
(15) Silverstein, H. J.; Cruz-Kan, K. C.; Hallas, A. M.; Zhou, H.; Donaberger, R. M.; Hernden, B. C.; Bieringer, M.; Choi, E. S.; Hwang, J. M.; Wills, A. S.; Wiebe, C. R. *Chem. Mater.* **2012**.
(16) Cromer, D. T.; Liberman, D. A. *Acta Crystallographica Section A* **1981**, *37*, 267–268.
(17) Rodríguez-Carvajal, J. *Newsletter of the Commission on Powder Diffraction* **2001**, *26*, 12–19.
(18) Rodríguez-Carvajal, J. *Physica B: Condensed Matter* **1993**, *192*, 55–69.
(19) Wills, A. S. *Physica B: Condensed Matter* **2000**, *276*, 680–681.
(20) West, D. V.; Davies, P. K. *J. Appl. Crystallogr.* **2011**, *44*, 595–602.
(21) Baur, W. H. *Acta Crystallographica Section B Structural Crystallography and Crystal Chemistry* **1974**, *30*, 1195–1215.
(22) Shannon, R. D.; Prewitt, C. T. *Acta Crystallographica Section B: Structural Crystallography and Crystal Chemistry* **1969**, *25*, 925–946.
(23) Kovalev *Representation of Crystallographic Space Groups: Irreducible Representations, Induced Representation and Corepresentations*; CRC Press, 1993.
(24) Mill, B. V.; Maksimov, B. A.; Pisarevsky, Y. V.; Danilova, N. P.; Markina, M. P.; Pavlovska, A.; Werner S; Schneider, J. In *Frequency Control Symposium and Exposition, 2004. Proceedings of the 2004 IEEE International*; IEEE, 2004; pp. 52– 60.
(25) Lam, A. E.; Groat, L. A.; Cooper, M. A.; Hawthorne, F. C. *The Canadian Mineralogist* **1994**, *32*, 525–532.
(26) Bérar, J. F.; Lelann, P. *Journal of Applied Crystallography* **1991**, *24*, 1–5.

**Figures**

Figure 1: The crystal structure of $Ba_3TeCo_3P_2O_{14}$, which has the prototypical $Ca_3Ga_2Ge_4O_{14}$ structure type. Each crystallographic site is differentiated by color (online). Shown are the individual layers that comprise the structure. In the lower right image, shaded triangles join the magnetic cations to in the magnetic sublattice and show the two different types of cobalt triangles present.

Figure 2: Rietveld refinement of synchrotron diffraction data for $Ba_3TeCo_3P_2O_{14}$ from 11-BM. Upper inset shows fit at high angles. Lower inset shows the peak asymmetry that was modeled by the inclusion of a second "dummy phase" in the refinement.

Figure 3: Combined Rietveld refinements of $Ba_3TeCo_3P_2O_{14}$ using the 2.41 Å and 1.54 Å NPD data. Inset highlights the magnetic peaks in the 2.41 Å diffraction data.

Figure 4: Rietveld refinements of synchrotron diffraction data from 11-BM. Top: $Pb_3TeCo_3V_2O_{14}$ Bottom: $Pb_3TeCo_3P_2O_{14}$ Upper insets highlight splitting of trigonal peaks that confirm the monoclinic distortion. Lower insets highlight the fit of the P2 supercell with six times increase in volume over the standard Langasite cell.

Figure 5: Combined Rietveld refinements of $Pb_3TeCo_3P_2O_{14}$ using the 2.41 Å and 1.54 Å NPD data. Inset highlights the magnetic peaks in the 2.41 Å diffraction data.

Figure 6: Crystal Structures of $Pb_3TeCo_3P_2O_{14}$ and $Ba_3TeCo_3P_2O_{14}$ as viewed along the c-axis. Arrows show direction of ion displacement in $Pb_3TeCo_3P_2O_{14}$ relative to the higher symmetry $Ba_3TeCo_3P_2O_{14}$. Also shown are the respective unit cells for spatial comparison (cells normalized for comparison).

Figure 7: Comparison of Pb-O and Ba-O coordination showing the effect of $Pb^{2+}$ stereochemical lone pairs. Two bond distances are given to account for the split oxygen position.

Figure 8: Rietveld refinement of the laboratory XRD data for $Pb_3TeZn_3P_2O_{14}$. Structural model is taken from $Pb_3TeCo_3P_2O_{14}$; the only changes made were the substitution of zinc for cobalt, the unit cell size, and the overall Debye-Waller factor. Similar peaks are found in $Pb_3TeZn_3V_2O_{14}$.

Figure 9: Temperature dependent magnetic susceptibility (upper panel) and inverse susceptibility (lower panel) for $Pb_3TeCo_3V_2O_{14}$, $Pb_3TeCo_3P_2O_{14}$, and $Ba_3TeCo_3P_2O_{14}$. Samples show distinct antiferromagnetic ordering and minimal frustration as evidenced by index of frustration ($f=|\theta_W|/T_N$). Inset of parameters determined by fitting the data on heating to the Curie-Weiss law.

Figure 10: Temperature dependent heat capacity normalized per mole of cobalt. $Pb_3TeCo_3V_2O_{14}$ is the only variant that exhibits two distinct magnetic transitions in zero field. The magnetic component of the heat capacity and the entropy release show that there is a significant amount of entropy released above $T_n$ in all cases. Vertical lines are drawn to illustrate transition temperatures across all panels. Inset shows heat capacity comparison of $Ba_3TeCo_3P_2O_{14}$ and nonmagnetic $Ba_3TeZn_3P_2O_{14}$.

Figure 11: Field dependent magnetization of compounds up to 16 T. Each illustrates field driven transitions suggesting that this is inherent to cobalt in this type of magnetic sublattice. $Ba_3TeCo_3P_2O_{14}$ and $Pb_3TeCo_3P_2O_{14}$ exhibit large hysteresis. Arrows indicate the direction of field change.

Figure 12: Magnetic structure models of $Ba_3TeCo_3P_2O_{14}$ and $Pb_3TeCo_3P_2O_{14}$. Both structures exhibit ferromagnetic coupling within each triangle as well as antiferromagnetic coupling between triangles and layers. Structures are shown from the same perspective for ease of comparison.

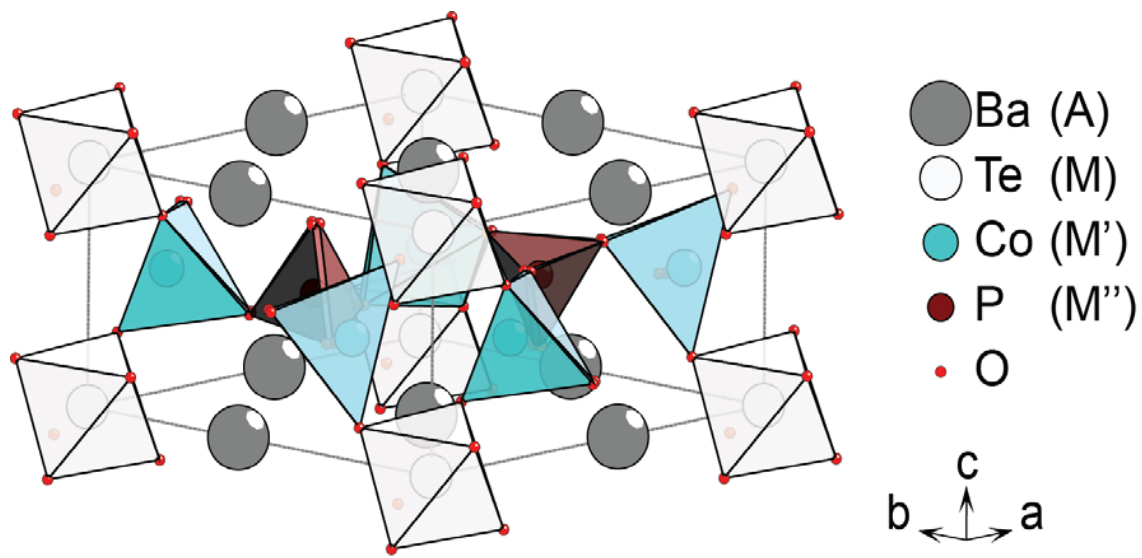
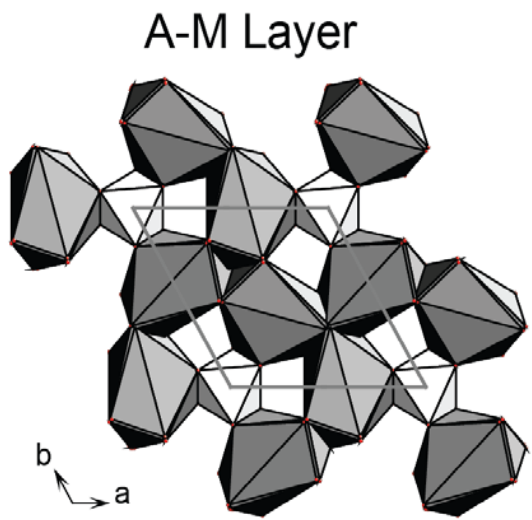
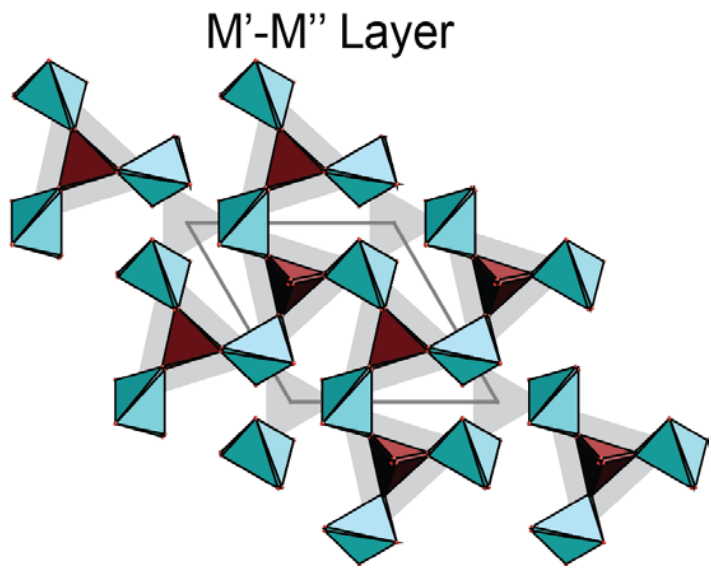

Figure 1:

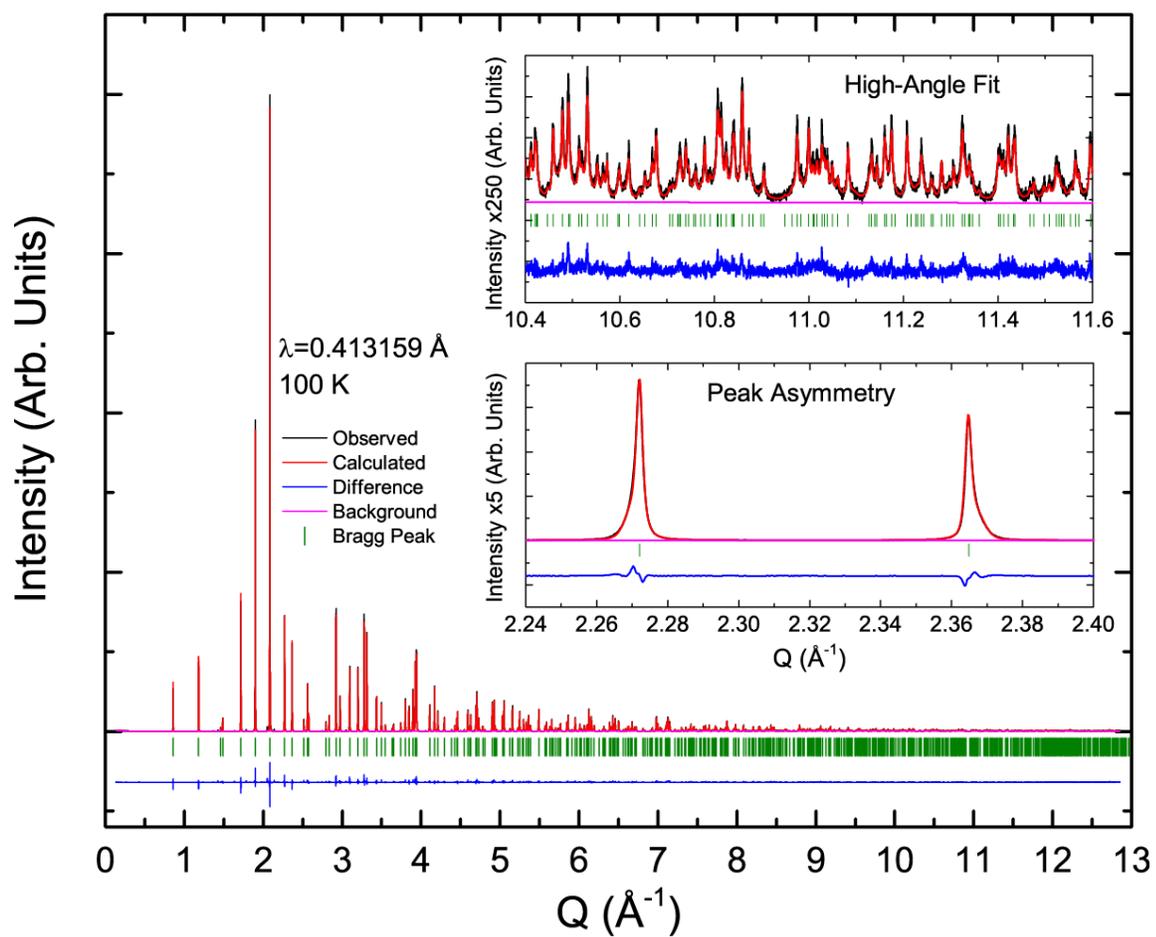

Figure 2:

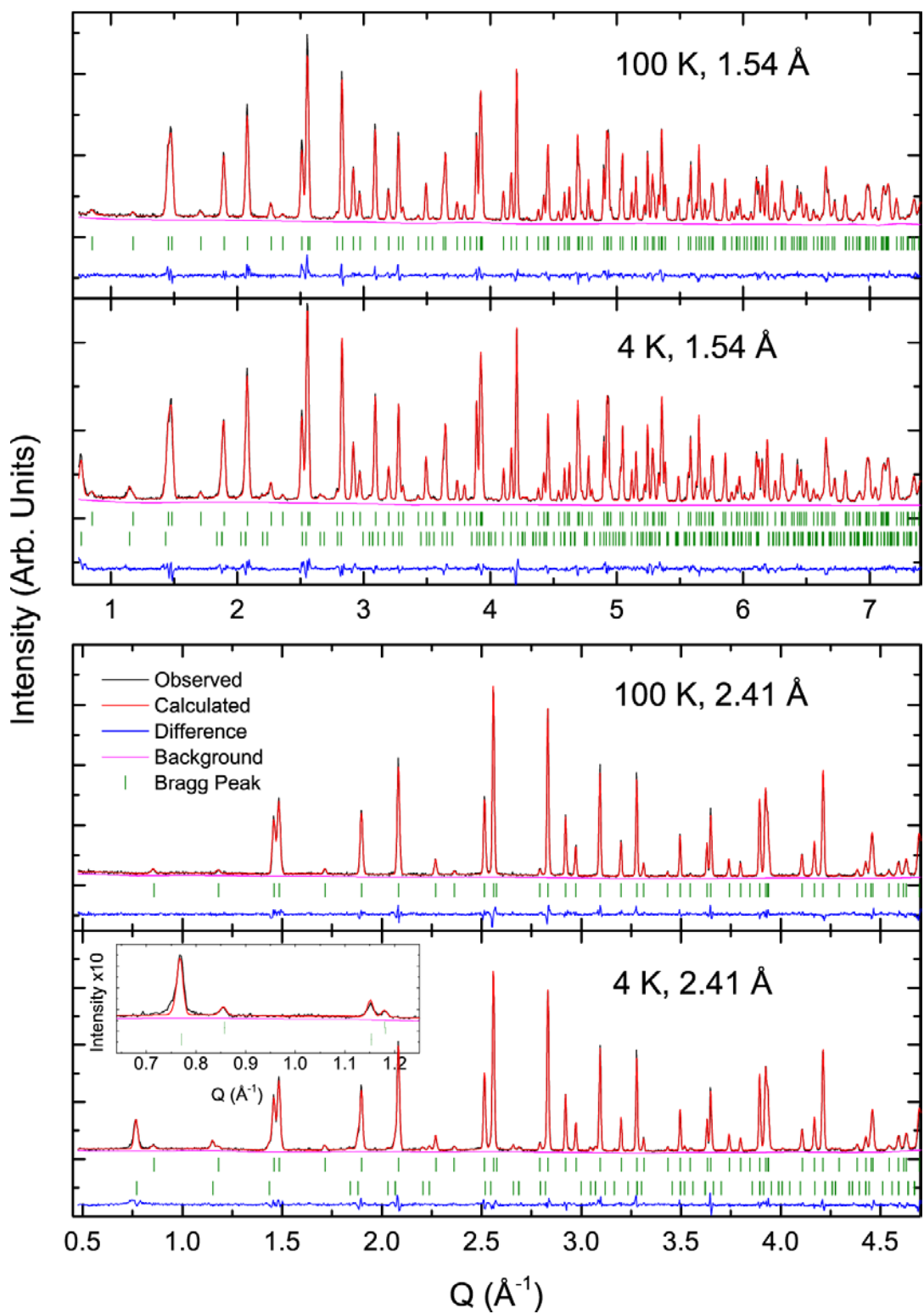

Figure 3:

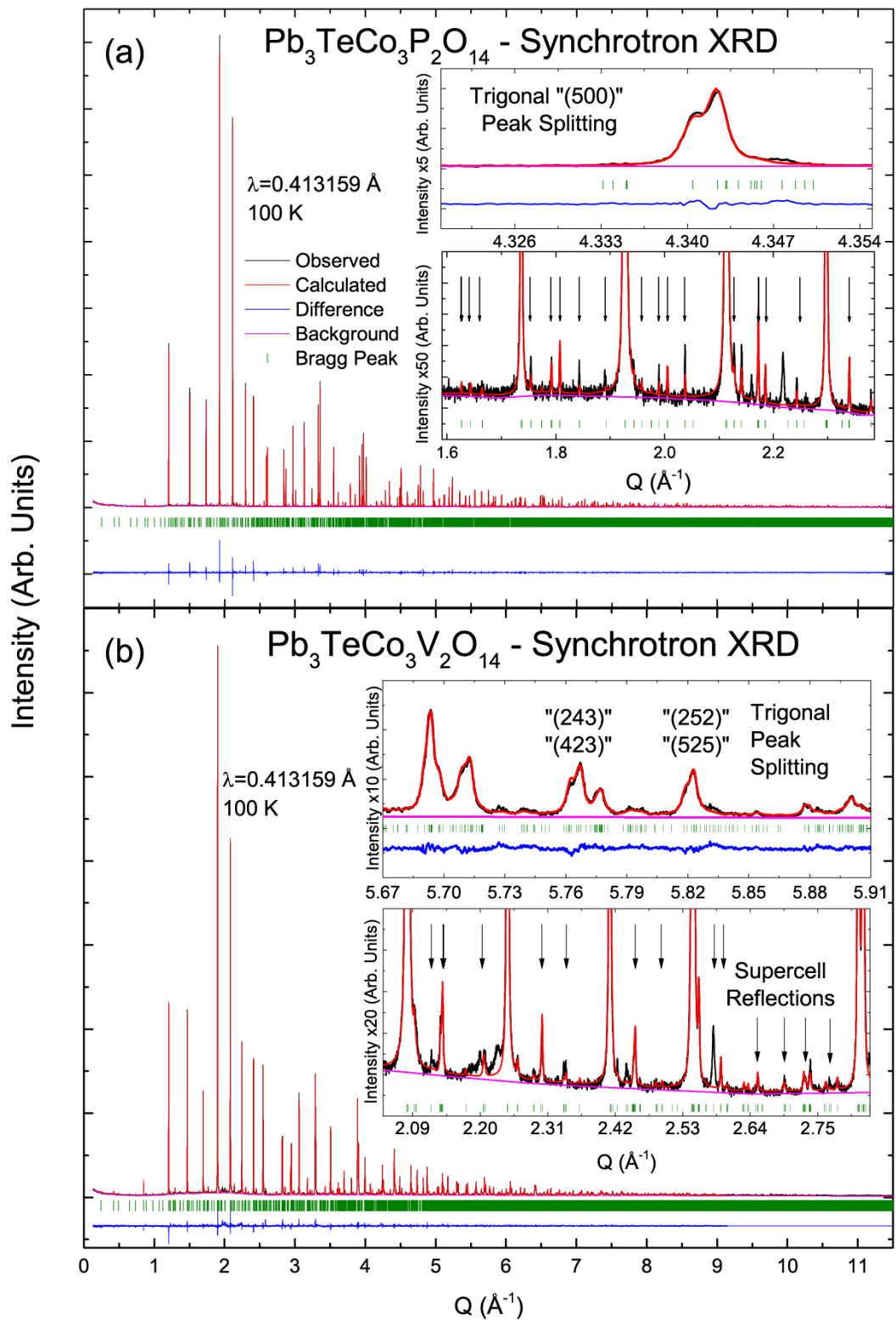

Figure 4:

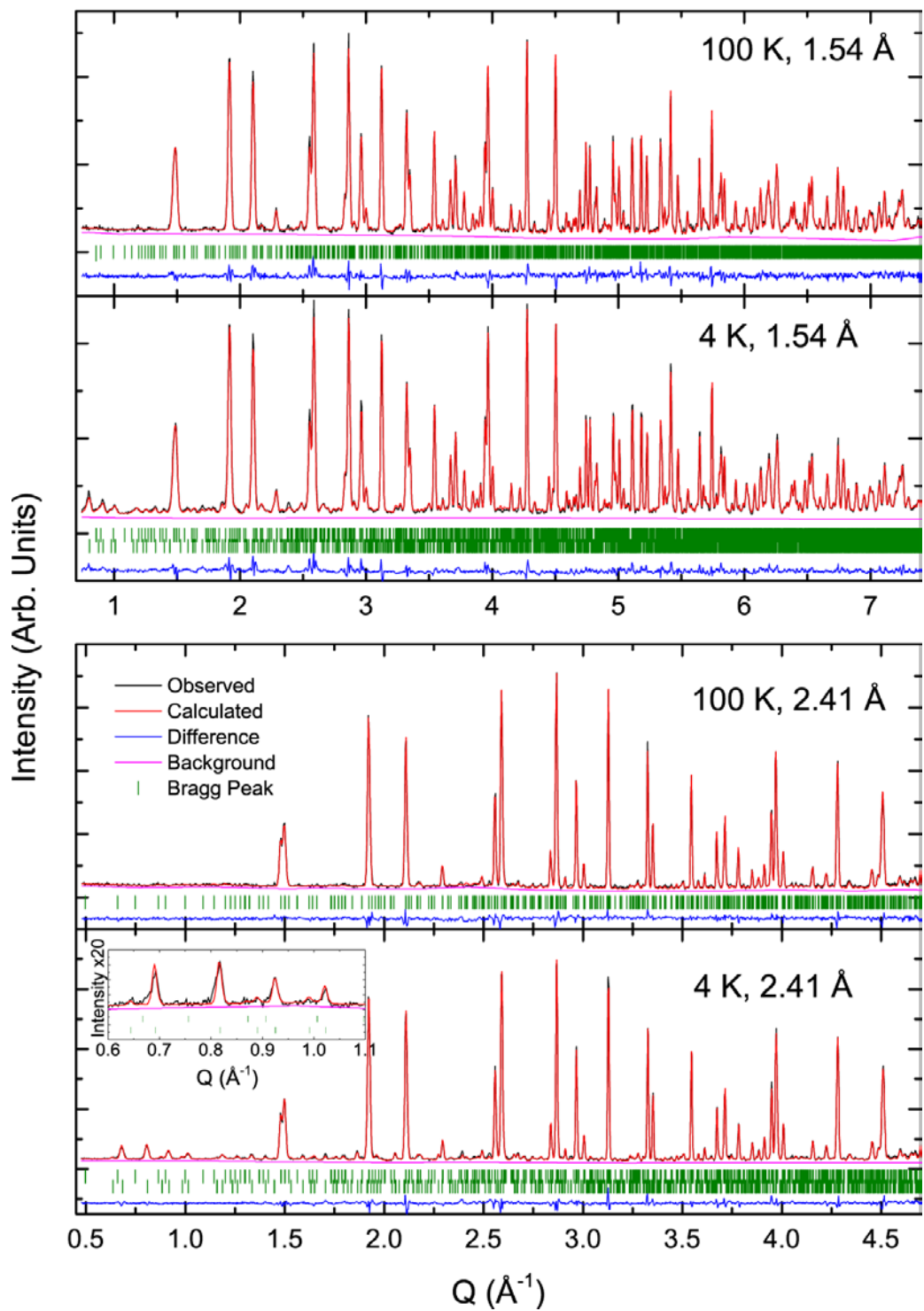

Figure 5:

Pb$_3$TeCo$_3$P$_2$O$_{14}$:
Comparison to Ba$_3$TeCo$_3$P$_2$O$_{14}$

P321

P2

- Pb
- Te
- Co
- P
- O

Figure 6:

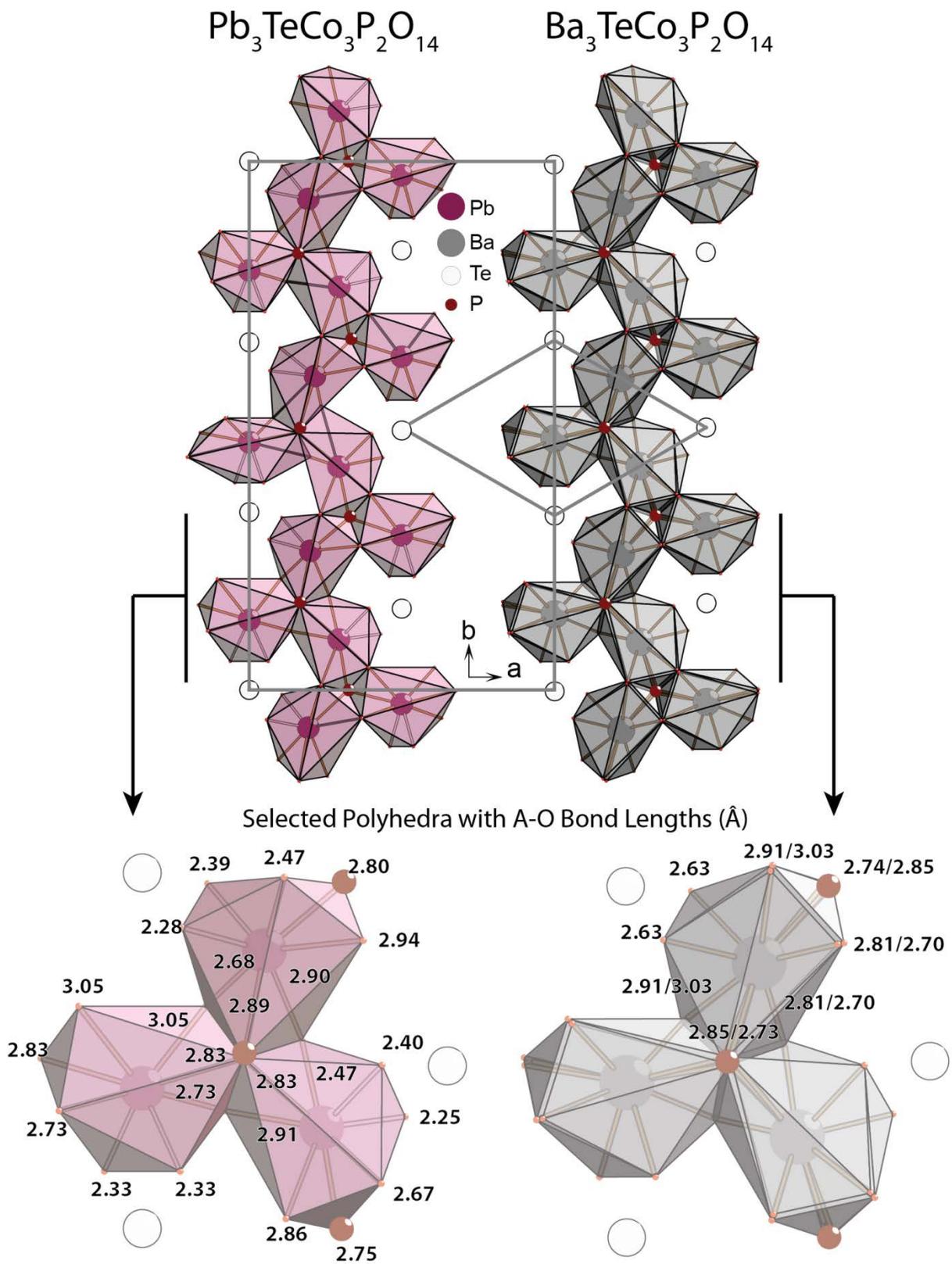

Figure 7:

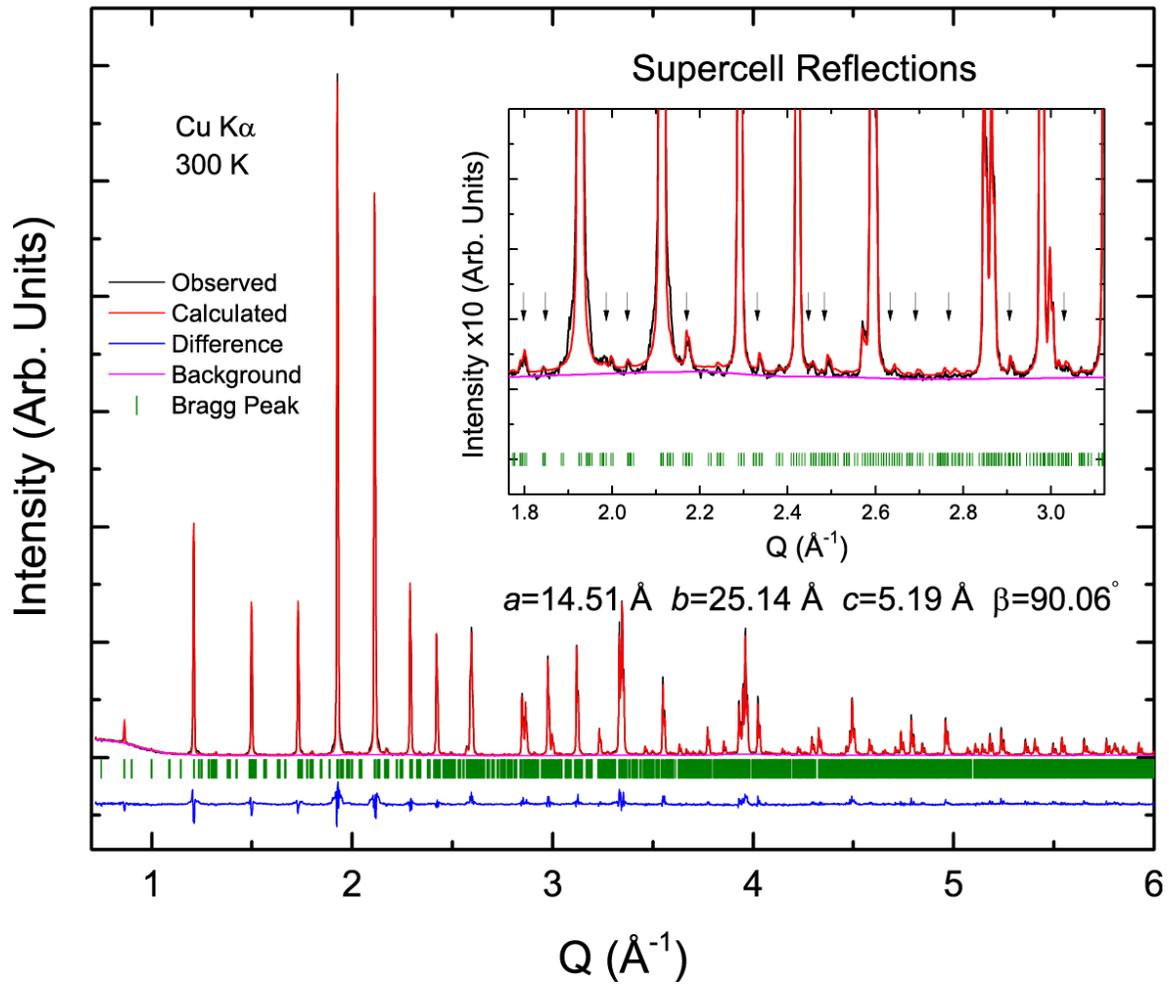

Figure 8:

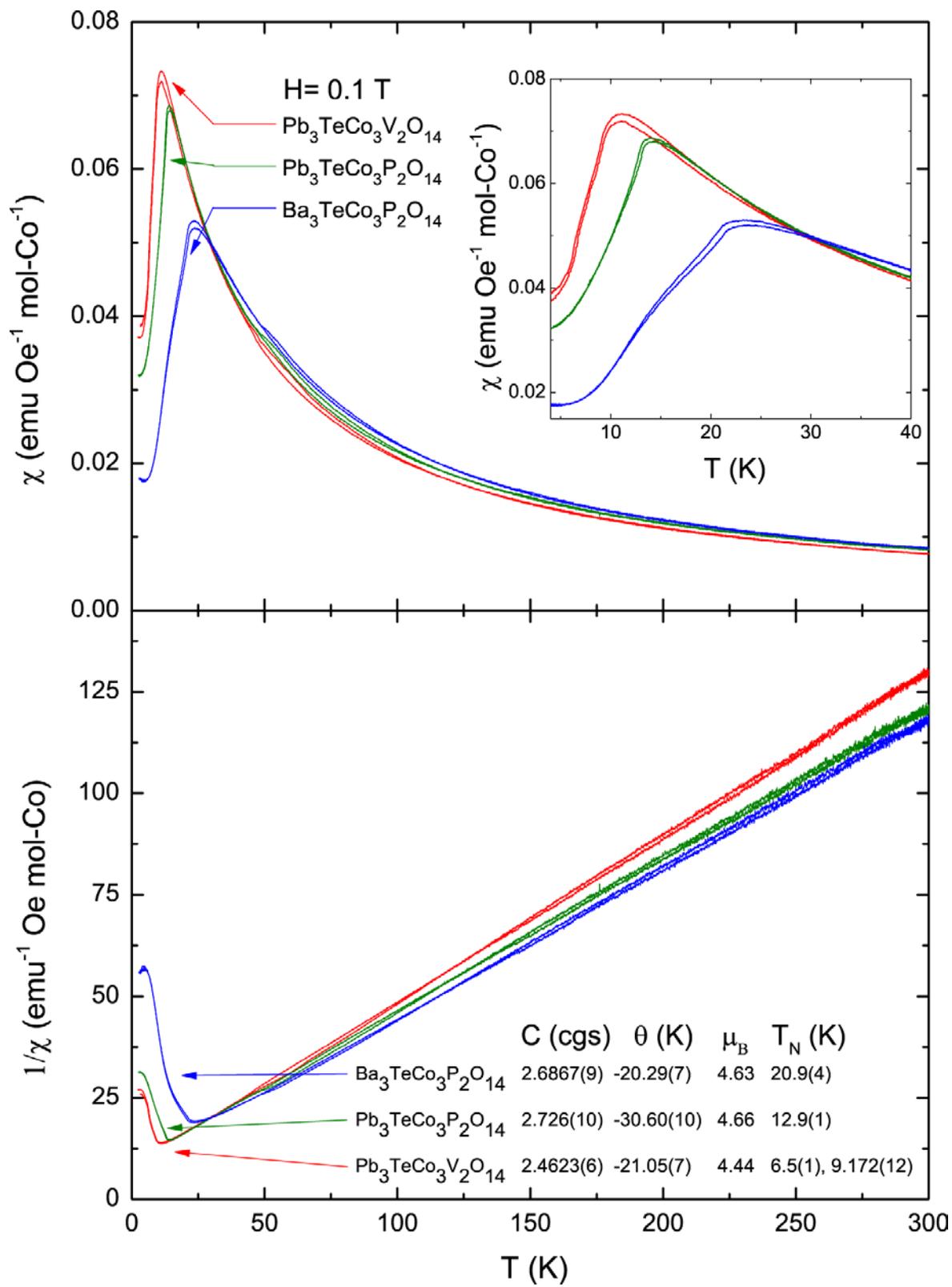

Figure 9:

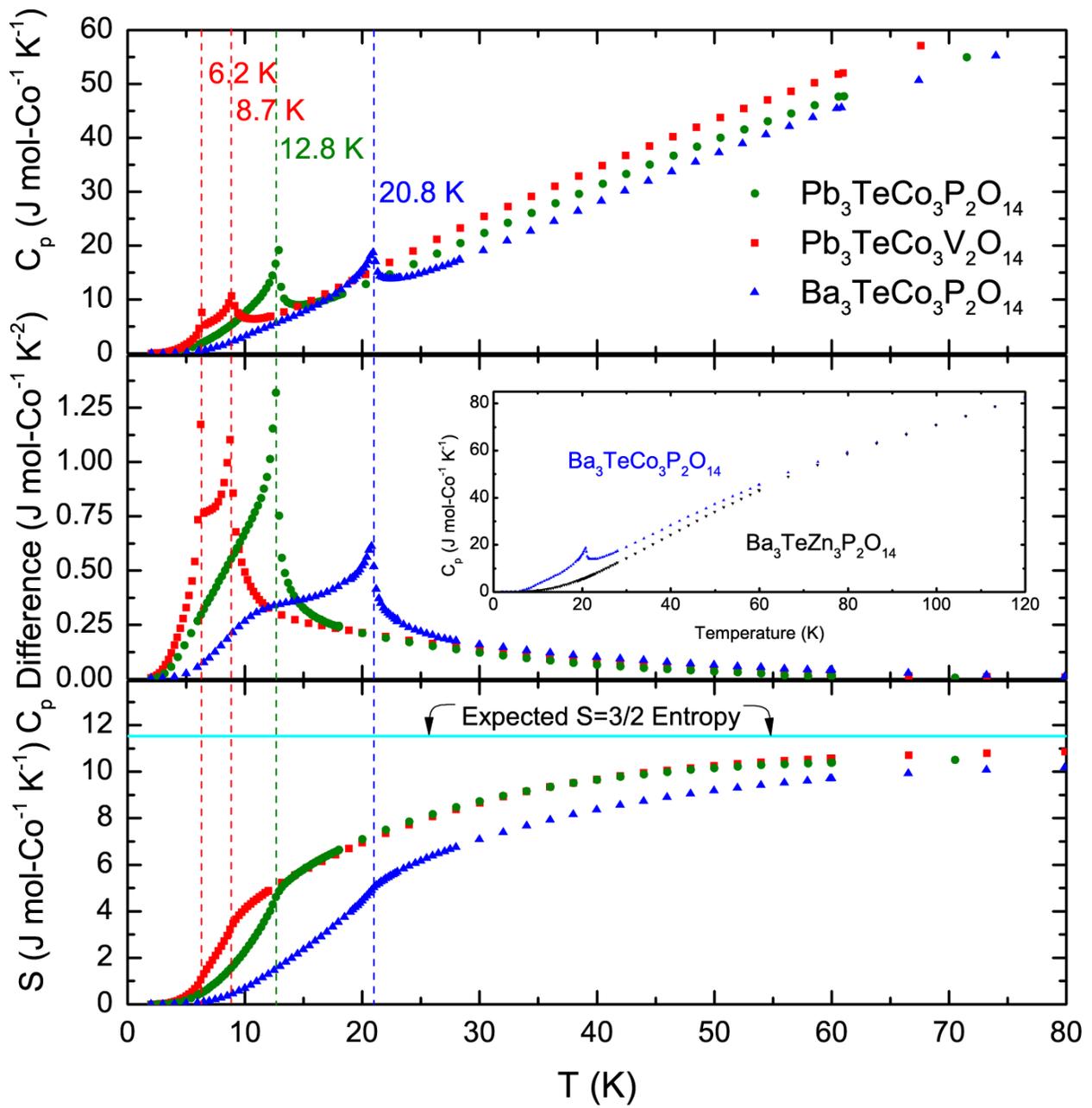

Figure 10:

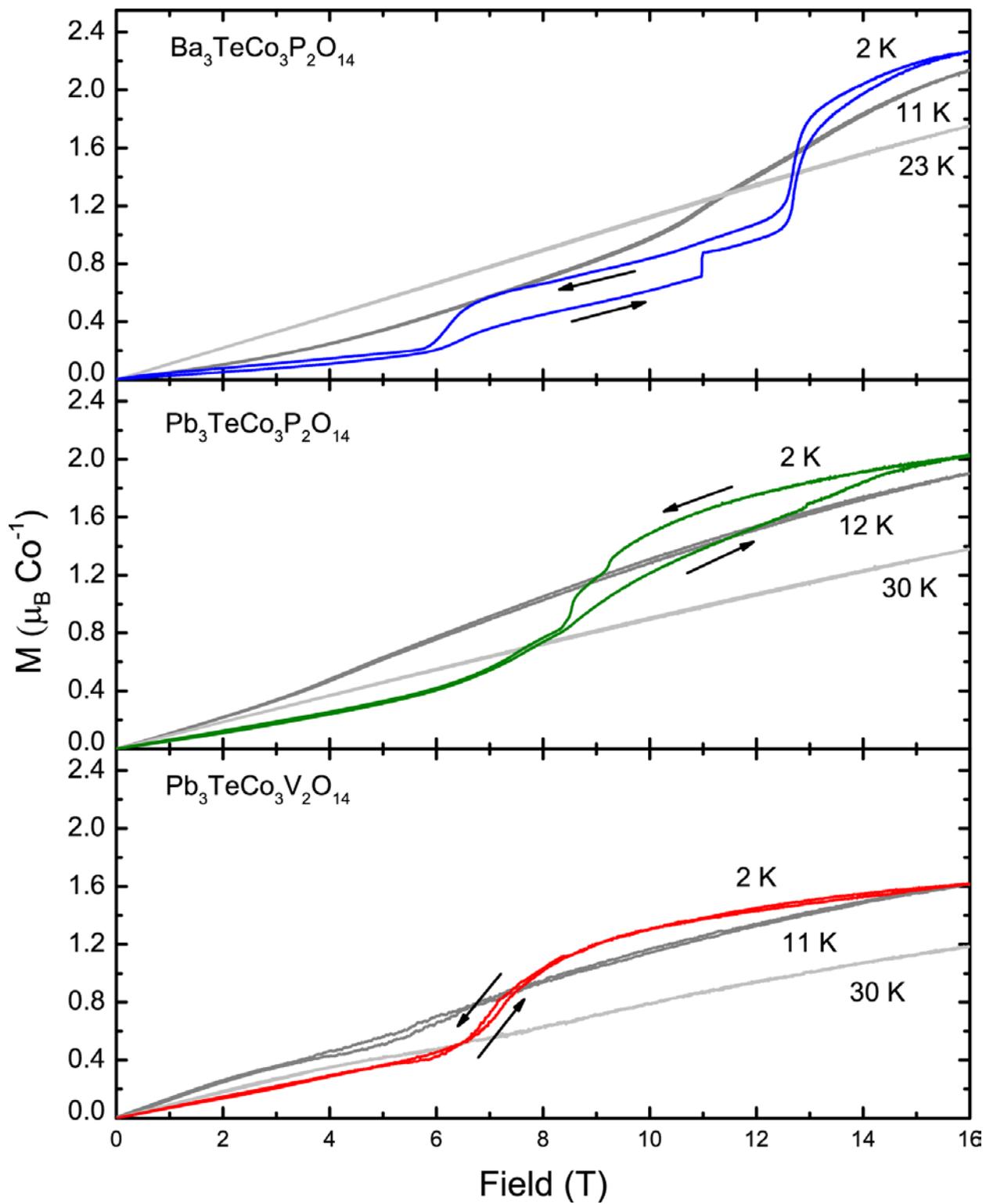

Figure 11:

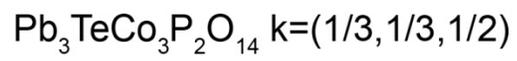
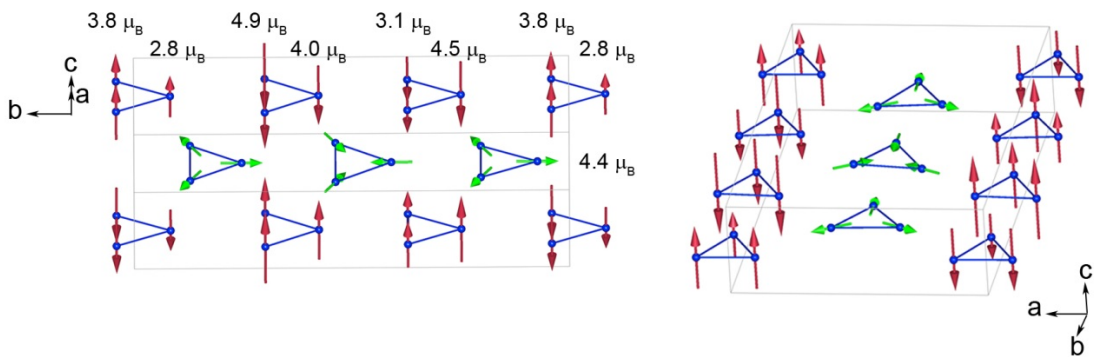
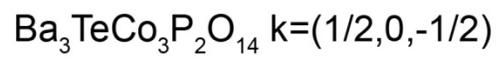
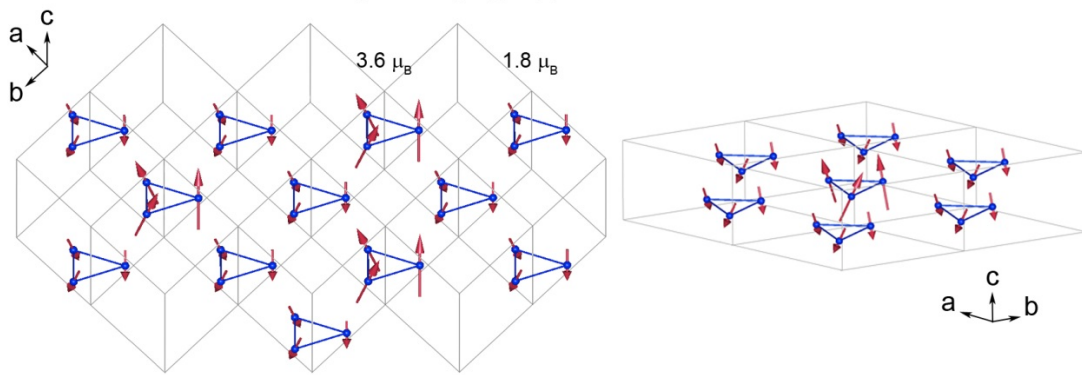

Figure 12:

Table 1: Crystal Structure of $Ba_3TeCo_3P_2O_{14}$[†]

| Primary Phase | | Space Group | | 100 K | a=8.44862(2) Å | | b=8.44862(2) Å | | c=5.313998(13) Å | |
|---|---|---|---|---|---|---|---|---|---|---|
| | | Z = 1 | | 4 K | a=8.4464(6) Å | | a=8.4464(6) Å | | c=5.3146(5) Å | |
| | | P 3 2 1 | | | α=90° | | β=90° | | γ=120° | |
| Atom | Site | X | y | z | Biso | U11 | U22 | U33 | U12 | U13 | U23 |
| Ba | 3e | 0.56161(5) | 0 | 0 | 0.243 | 0.0013(2) | 0.0009(3) | 0.0020(4) | 0.00045(13) | 0.0000(5) | 0.000(1) |
| Te | 1a | 0 | 0 | 0 | 0.287 | 0.0016(5) | 0.0016(5) | 0.002(1) | 0.008(3) | 0.00 | 0.00 |
| Co | 3f | 0.23437(5) | 0 | 1/2 | 0.1934(5) | | | | | | |
| P | 2d | 2/3 | 1/3 | 0.53718(5) | 0.184(2) | | | | | | |
| O1* | 6g | 0.6602(6) | 0.3415(5) | 0.25059(5) | 0.065(6) | | | | | | |
| O2** | 6g | 0.5227(1) | 0.71845(5) | 0.3516(1) | 0.200(2) | Occ. = 0.5 | | | | | |
| O3** | 6g | 0.51686(5) | 0.69851(5) | 0.3522(1) | 0.200(2) | Occ. = 0.5 | | | | | |
| O4 | 6g | 0.78546(5) | 0.90707(5) | 0.79196(5) | 0.240 | 0.0015(16) | 0.001(2) | 0.000(2) | 0.000(1) | 0.00 | 0.00 |

† Estimated standard deviations (ESDs) were multiplied by 5.19 to account for the presence of correlated residuals.[26]

\* Apical Oxygen
\*\* Base Oxygen

Table 2: Soft constraints implemented in the refinement of $Pb_3TeCo_3P_2O_{14}$

| Bond | Length (Å) | Sigma (Å) |
|---|---|---|
| Co-O | 1.97 | 0.09 |
| Ta-O | 1.93 | 0.05 |
| P-O* | 1.531 | 0.01 |
| P-O** | 1.546 | 0.01 |

\* Apical Oxygen
\*\* Base Oxygen

Table 3: Crystal Structure of Pb$_3$TeCo$_3$P$_2$O$_{14}$†

| Monoclinic | | | | 100 K | a=14.47600(6) Å | | b=25.0579(1) Å | | c=5.214222(16) Å | |
|---|---|---|---|---|---|---|---|---|---|---|
| | | | | 4 K | a=14.47371(5) Å | | b=25.0378(4) Å | | c=5.21340(9) Å | |
| Space Group P 1 2 1 | | Z = 6 | | | α=90° | | β=90.0304(4) ° | | γ=90° | |
| Type | Site | x | y | z | B$_{iso}$ | Type | Type | x | y | z | B$_{iso}$ |
| Pb1 | 1a | 0 | 0.4686(17) | 0 | 0.289(3) | Co4 | 1d | 1/2 | 0.409(4) | 1/2 | 0.17(1) |
| Pb2 | 1c | 1/2 | 0.9734(17) | 0 | 0.289(3) | Co5 | 1b | 0 | 0.248(4) | 1/2 | 0.17(1) |
| Pb3 | 1a | 0 | 0.794(2) | 0 | 0.289(3) | Co6 | 1d | 1/2 | 0.752(4) | 1/2 | 0.17(1) |
| Pb4 | 1c | 1/2 | 0.292(2) | 0 | 0.289(3) | Co7 | 2e | 0.114(5) | 0.041(3) | 0.497(18) | 0.17(1) |
| Pb5 | 1a | 0 | 0.133(2) | 0 | 0.289(3) | Co8 | 2e | 0.632(5) | 0.537(3) | 0.523(17) | 0.17(1) |
| Pb6 | 1c | 1/2 | 0.631(2) | 0 | 0.289(3) | Co9 | 2e | 0.119(5) | 0.699(3) | 0.510(18) | 0.17(1) |
| Pb7 | 2e | 0.2941(15) | 0.0953(15) | 0.990(5) | 0.289(3) | Co10 | 2e | 0.626(5) | 0.203(3) | 0.485(5) | 0.17(1) |
| Pb8 | 2e | 0.7848(15) | 0.5940(15) | 0.003(5) | 0.289(3) | Co11 | 2e | 0.120(5) | 0.371(3) | 0.488(18) | 0.17(1) |
| Pb9 | 2e | 0.295(3) | 0.7627(14) | 0.009(7) | 0.289(3) | Co12 | 2e | 0.634(5) | 0.871(3) | 0.493(19) | 0.17(1) |
| Pb10 | 2e | 0.800(2) | 0.2630(14) | 0.987(7) | 0.289(3) | O19* | 2e | 0.166(9) | 0.161(6) | 0.253(5) | 0.256(1) |
| Pb11 | 2e | 0.293(2) | 0.4225(13) | 0.013(6) | 0.289(3) | P1 | 2e | 0.1676(17) | 0.166(9) | 0.546(4) | 0.256(1) |
| Pb12 | 2e | 0.806(2) | 0.9288(13) | -0.002(7) | 0.289(3) | O20** | 2e | 0.136(8) | 0.111(2) | 0.65(2) | 0.256(1) |
| O1 | 2e | 0.050(8) | 0.716(4) | 0.21(2) | 0.17(1) | O21** | 2e | 0.104(7) | 0.210(4) | 0.65(2) | 0.256(1) |
| Te1 | 1a | 0 | 0.658(2) | 0 | 0.17(1) | O22** | 2e | 0.262(5) | 0.178(5) | 0.67(2) | 0.256(1) |
| O2 | 2e | 0.899(7) | 0.648(5) | 0.23(2) | 0.17(1) | O23* | 2e | 0.332(9) | 0.662(6) | 0.737(5) | 0.256(1) |
| O3 | 2e | 0.068(7) | 0.604(4) | 0.20(2) | 0.17(1) | P2 | 2e | 0.3351(17) | 0.664(9) | 0.443(4) | 0.256(1) |
| O4 | 2e | 0.541(9) | 0.215(4) | 0.21(2) | 0.17(1) | O24** | 2e | 0.393(7) | 0.712(3) | 0.35(2) | 0.256(1) |
| Te2 | 1c | 1/2 | 0.154(2) | 0 | 0.17(1) | O25** | 2e | 0.237(4) | 0.670(5) | 0.33(2) | 0.256(1) |
| O5 | 2e | 0.393(8) | 0.154(5) | 0.22(20) | 0.17(1) | O26** | 2e | 0.377(8) | 0.612(3) | 0.34(2) | 0.256(1) |
| O6 | 2e | 0.569(8) | 0.106(5) | 0.21(20) | 0.17(1) | O27* | 2e | 0.163(4) | 0.828(3) | 0.246(4) | 0.256(1) |
| O7 | 2e | 0.062(7) | 0.054(4) | 0.167(18) | 0.17(1) | P3 | 2e | 0.1608(16) | 0.8261(9) | 0.540(4) | 0.256(1) |
| Te3 | 1a | 0 | 0 | 0 | 0.17(1) | O28** | 2e | 0.095(7) | 0.868(3) | 0.66(3) | 0.256(1) |
| O8 | 2e | 0.057(6) | 0.944(4) | 0.19(20) | 0.17(1) | O29** | 2e | 0.256(4) | 0.837(5) | 0.67(2) | 0.256(1) |
| O9 | 2e | 0.892(7) | 0.992(5) | 0.21(2) | 0.17(1) | O30** | 2e | 0.122(8) | 0.773(3) | 0.64(2) | 0.256(1) |
| O10 | 2e | 0.548(8) | 0.548(4) | 0.22(2) | 0.17(1) | O31* | 2e | 0.328(8) | 0.325(6) | 0.744(5) | 0.256(1) |
| Te4 | 1c | 1/2 | 0.491(2) | 0 | 0.17(1) | P4 | 2e | 0.3314(16) | 0.3277(10) | 0.450(4) | 0.256(1) |
| O11 | 2e | 0.402(7) | 0.485(5) | 0.23(2) | 0.17(1) | O32** | 2e | 0.397(7) | 0.371(4) | 0.34(2) | 0.256(1) |
| O12 | 2e | 0.562(7) | 0.440(4) | 0.22(2) | 0.17(1) | O33** | 2e | 0.232(4) | 0.332(5) | 0.34(2) | 0.256(1) |
| O13 | 2e | 0.047(8) | 0.392(4) | 0.22(2) | 0.17(1) | O34** | 2e | 0.363(9) | 0.273(3) | 0.35(3) | 0.256(1) |
| Te5 | 1a | 0 | 0.336(2) | 0 | 0.17(1) | O35* | 2e | 0.160(9) | 0.487(4) | 0.211(6) | 0.256(1) |
| O14 | 2e | 0.893(7) | 0.326(5) | 0.21(2) | 0.17(1) | P5 | 2e | 0.1667(17) | 0.4960(9) | 0.500(4) | 0.256(1) |
| O15 | 2e | 0.065(7) | 0.285(4) | 0.21(2) | 0.17(1) | O36** | 2e | 0.078(3) | 0.515(4) | 0.637(14) | 0.256(1) |
| O16 | 2e | 0.539(8) | 0.893(4) | 0.21(2) | 0.17(1) | O37** | 2e | 0.235(5) | 0.539(3) | 0.59(2) | 0.256(1) |
| Te6 | 1c | 1/2 | 0.832(2) | 0 | 0.17(1) | O38** | 2e | 0.202(6) | 0.4402(18) | 0.577(18) | 0.256(1) |
| O17 | 2e | 0.397(7) | 0.827(5) | 0.22(2) | 0.17(1) | O39* | 2e | 0.322(8) | 0.998(6) | 0.754(5) | 0.256(1) |
| O18 | 2e | 0.567(8) | 0.780(4) | 0.20(2) | 0.17(1) | P6 | 2e | 0.3273(17) | 0.9999(10) | 0.460(4) | 0.256(1) |
| Co1 | 1b | 0 | 0.575(4) | 1/2 | 0.17(1) | O40** | 2e | 0.392(7) | 0.043(4) | 0.34(2) | 0.256(1) |
| Co2 | 1d | 1/2 | 0.073(4) | 1/2 | 0.17(1) | O41** | 2e | 0.236(4) | 0.009(5) | 0.312(18) | 0.256(1) |
| Co3 | 1b | 0 | 0.916(5) | 1/2 | 0.17(1) | O42** | 2e | 0.369(8) | 0.947(3) | 0.36(2) | 0.256(1) |

† ESDs were multiplied by 2.79 to account for the presence of correlated residuals.[26]

\* Apical Oxygen
\*\* Base Oxygen

Table 4: Crystal Structure of $Pb_3TeCo_3V_2O_{14}$[†]

| Monoclinic | | | | | | a=14.81359(9) Å | | b=25.64683(16) Å | | c=5.21137(3) | |
|---|---|---|---|---|---|---|---|---|---|---|---|
| Space Group P 1 2 1 | | | Z = 6 | | | α=90° | | β=90.0631(8)° | | γ=90° | |
| Type | Site | x | y | z | $B_{iso}$‡ | Type | Site | x | y | z | $B_{iso}$‡ |
| Pb1 | 1a | 0 | 0.466(2) | 0 | 0.55(2) | Co4 | 1d | 1/2 | 0.416(4) | 1/2 | 0.3 |
| Pb2 | 1c | 1/2 | 0.972(2) | 0 | 0.55(2) | Co5 | 1b | 0 | 0.254(4) | 1/2 | 0.3 |
| Pb3 | 1a | 0 | 0.790(2) | 0 | 0.55(2) | Co6 | 1d | 1/2 | 0.756(4) | 1/2 | 0.3 |
| Pb4 | 1c | 1/2 | 0.287(2) | 0 | 0.55(2) | Co7 | 2e | 0.137(4) | 0.044(3) | 0.48(2) | 0.3 |
| Pb5 | 1a | 0 | 0.130(2) | 0 | 0.55(2) | Co8 | 2e | 0.617(5) | 0.534(3) | 0.52(2) | 0.3 |
| Pb6 | 1c | 1/2 | 0.626(2) | 0 | 0.55(2) | Co9 | 2e | 0.113(5) | 0.696(3) | 0.51(2) | 0.3 |
| Pb7 | 2e | 0.296(1) | 0.096(2) | -0.009(5) | 0.55(2) | Co10 | 2e | 0.620(5) | 0.200(3) | 0.51(2) | 0.3 |
| Pb8 | 2e | 0.788(1) | 0.594(2) | 0.005(5) | 0.55(2) | Co11 | 2e | 0.114(5) | 0.378(3) | 0.48(2) | 0.3 |
| Pb9 | 2e | 0.300(2) | 0.764(2) | -0.001(6) | 0.55(2) | Co12 | 2e | 0.630(5) | 0.867(3) | 0.50(2) | 0.3 |
| Pb10 | 2e | 0.807(2) | 0.265(2) | -0.009(6) | 0.55(2) | O19 | 2e | 0.16702 | 0.15831 | 0.19745 | 0.15 |
| Pb11 | 2e | 0.298(1) | 0.419(2) | 0.022(4) | 0.55(2) | V1 | 2e | 0.17(1) | 0.163(7) | 0.53(3) | 0.3 |
| Pb12 | 2e | 0.818(1) | 0.930(2) | -0.001(5) | 0.55(2) | O20 | 2e | 0.13098 | 0.10437 | 0.67756 | 0.15 |
| O1 | 2e | 0.04424 | 0.71084 | 0.2124 | 0.2 | O21 | 2e | 0.09285 | 0.21116 | 0.64382 | 0.15 |
| Te1 | 1a | 0 | 0.655(5) | 0 | 0.4 | O22 | 2e | 0.27223 | 0.17712 | 0.6599 | 0.15 |
| O2 | 2e | -0.1059 | 0.649 | 0.2124 | 0.2 | O23 | 2e | 0.33201 | 0.66788 | 0.77637 | 0.15 |
| O3 | 2e | 0.06191 | 0.60482 | 0.2124 | 0.2 | V2 | 2e | 0.33(1) | 0.667(7) | 0.44(3) | 0.3 |
| O4 | 2e | 0.53762 | 0.20884 | 0.2124 | 0.2 | O24 | 2e | 0.39797 | 0.71829 | 0.31304 | 0.15 |
| Te2 | 1c | 1/2 | 0.151(5) | 0 | 0.4 | O25 | 2e | 0.22724 | 0.67103 | 0.3042 | 0.15 |
| O5 | 2e | 0.39503 | 0.1413 | 0.2124 | 0.2 | O26 | 2e | 0.38254 | 0.60926 | 0.32078 | 0.15 |
| O6 | 2e | 0.5676 | 0.10374 | 0.2124 | 0.2 | O27 | 2e | 0.16047 | 0.83035 | 0.23667 | 0.15 |
| O7 | 2e | 0.04338 | 0.05618 | 0.2124 | 0.2 | V3 | 2e | 0.16(1) | 0.827(7) | 0.57(3) | 0.3 |
| Te3 | 1a | 0 | 0 | 0 | 0.4 | O28 | 2e | 0.08486 | 0.87354 | 0.70568 | 0.15 |
| O8 | 2e | 0.06267 | -0.04975 | 0.2124 | 0.2 | O29 | 2e | 0.26343 | 0.83764 | 0.71094 | 0.15 |
| O9 | 2e | -0.1058 | -0.00642 | 0.2124 | 0.2 | O30 | 2e | 0.12095 | 0.76622 | 0.68305 | 0.15 |
| O10 | 2e | 0.55269 | 0.54555 | 0.2124 | 0.2 | O31 | 2e | 0.32659 | 0.32624 | 0.80969 | 0.15 |
| Te4 | 1c | 1/2 | 0.492(5) | 0 | 0.4 | V4 | 2e | 0.32(1) | 0.329(7) | 0.47414 | 0.3 |
| O11 | 2e | 0.39362 | 0.49168 | 0.2124 | 0.2 | O32 | 2e | 0.39258 | 0.37835 | 0.35097 | 0.15 |
| O12 | 2e | 0.55395 | 0.43905 | 0.2124 | 0.2 | O33 | 2e | 0.21577 | 0.33966 | 0.3533 | 0.15 |
| O13 | 2e | 0.04189 | 0.39516 | 0.2124 | 0.2 | O34 | 2e | 0.3619 | 0.2705 | 0.33414 | 0.15 |
| Te5 | 1a | 0 | 0.339(5) | 0 | 0.4 | O35 | 2e | 0.14618 | 0.49552 | 0.15067 | 0.15 |
| O14 | 2e | -0.10562 | 0.33124 | 0.2124 | 0.2 | V5 | 2e | 0.155(9) | 0.50(3) | 0.49(2) | 0.3 |
| O15 | 2e | 0.06183 | 0.29107 | 0.20526 | 0.2 | O36 | 2e | 0.05265 | 0.51348 | 0.63317 | 0.15 |
| O16 | 2e | 0.54418 | 0.89283 | 0.2124 | 0.2 | O37 | 2e | 0.23405 | 0.54311 | 0.59176 | 0.15 |
| Te6 | 1c | 1/2 | 0.837(5) | 0 | 0.4 | O38 | 2e | 0.18852 | 0.43732 | 0.61456 | 0.15 |
| O17 | 2e | 0.39411 | 0.83094 | 0.2124 | 0.2 | O39 | 2e | 0.30838 | 0.98915 | 0.79302 | 0.15 |
| O18 | 2e | 0.56197 | 0.78682 | 0.2124 | 0.2 | V6 | 2e | 0.32(1) | 0.995(7) | 0.46(3) | 0.3 |
| Co1 | 1b | 0 | 0.572(4) | 1/2 | 0.3 | O40 | 2e | 0.39234 | 1.04513 | 0.37054 | 0.15 |
| Co2 | 1d | 1/2 | 0.079(4) | 1/2 | 0.3 | O41 | 2e | 0.21624 | 1.00683 | 0.30754 | 0.15 |
| Co3 | 1b | 0 | 0.923(4) | 1/2 | 0.3 | O42 | 2e | 0.36186 | 0.93766 | 0.31964 | 0.15 |

† ESDs were multiplied by 5.14 to account for the presence of correlated residuals.[26] $TeO_6$ and $VO_4$ polyhedra are refined as rigid body groups. ESDs are given for the refined position of the central atom only. Only SXRD refined.

‡ $B_{iso}$ estimated and fixed at chemically reasonable values for the light elements.

Table 5: Bond Lengths Compared – $Ba_3TeCo_3P_2O_{14}$ and $Pb_3TeCo_3P_2O_{14}$:

| $Ba_3TeCo_3P_2O_{14}$ | | |
|---|---|---|
| | Average (Å) | Range (Å) |
| Ba1-O1 | 2.83 | 2.62-3.02 |
| Co1-O1 | 2.00 | 1.91-2.04 |
| **$Pb_3TeCo_3P_2O_{14}$** | | |
| | Average (Å) | Range (Å) |
| Pb1-O | 2.79 | 2.33-3.73 |
| Pb2-O | 2.75 | 2.35-2.95 |
| Pb3-O | 2.68 | 2.36-2.89 |
| Pb4-O | 2.75 | 2.30-3.03 |
| Pb5-O | 2.74 | 2.33-3.05 |
| Pb6-O | 2.77 | 2.47-3.14 |
| Pb7-O | 2.67 | 2.25-2.91 |
| Pb8-O | 2.75 | 2.38-3.37 |
| Pb9-O | 2.73 | 2.32-3.17 |
| Pb10-O | 2.67 | 2.28-2.94 |
| Pb11-O | 2.81 | 2.46-3.75 |
| Pb12-O | 2.73 | 2.25-3.18 |
| Co1-O | 2.00 | 1.98-2.02 |
| Co2-O | 1.96 | 1.92-2.01 |
| Co3-O | 1.98 | 1.93-2.02 |
| Co4-O | 1.92 | 1.88-1.96 |
| Co5-O | 1.97 | 1.95-1.99 |
| Co6-O | 1.98 | 1.95-2.01 |
| Co7-O | 2.00 | 1.91-2.15 |
| Co8-O | 1.98 | 1.90-2.02 |
| Co9-O | 1.96 | 1.89-2.08 |
| Co10-O | 1.94 | 1.92-1.97 |
| Co11-O | 1.99 | 1.84-2.15 |
| Co12-O | 2.01 | 1.91-2.11 |

Table 6: R Values for combined Rietveld refinements

| | $Ba_3TeCo_3P_2O_{14}$ | | | $Pb_3TeCo_3P_2O_{14}$ | | | $Pb_3TeCo_3V_2O_{14}$ | | | $Pb_3TeZn_3P_2O_{14}$ | | |
|---|---|---|---|---|---|---|---|---|---|---|---|---|
| **Pattern** | $R_p$ | $R_{wp}$ | $\chi^2$ | $R_p$ | $R_{wp}$ | $\chi^2$ | $R_p$ | $R_{wp}$ | $\chi^2$ | $R_p$ | $R_{wp}$ | $\chi^2$ |
| 100K, 0.41 Å | 7.86 | 10.4 | 5.05 | 12.8 | 12.0 | 1.57 | 13.7 | 15 | 2.51 | | | |
| 100 K, 1.54 Å | 10.0 | 9.85 | 3.49 | 10.2 | 10.8 | 6.80 | | | | | | |
| 100 K, 2.41 Å | 14.3 | 12.5 | 3.52 | 13.8 | 13.8 | 4.39 | | | | | | |
| 4 K, 1.54 Å | 8.98 | 9.54 | 3.61 | 7.79 | 8.76 | 9.18 | | | | | | |
| 4 K, 2.41 Å | 13.7 | 12.9 | 6.06 | 10.4 | 11.1 | 9.64 | | | | | | |
| 300 K, Cu Kα | | | | | | | | | | 13.5 | 17.6 | 6.01 |

Table 7: Magnetic Structure of $Ba_3TeCo_3P_2O_{14}$

| Irreducible Representation: $\Gamma_1$ | k=(1/3, 1/3, 1/2) | | | |
|---|---|---|---|---|
| $Co_1$ | $\Psi_1$ | $\Psi_2$ | $C_1$ | $C_2$ |
| (x,y,z) | (2 0 0)+$i$(0 0 0) | (0 0 2)+$i$(0 0 0) | | |
| (-y, x-y, z) | (-2 -1 0)+$i$(0 0 0) | (0 0 2)+$i$(0 0 0) | 0.856 | 1.660 |
| (-x+y+1, -x+1, z) | (-0.5 -0.5 0)+$i$(0.866 -0.866 0) | (0 0 -1)+$i$(0 0 1.732) | | |

Two types of cobalt clusters exist in the structure; those which are comprised of 3.64 Bohr magneton moments and opposing clusters of 1.82 Bohr magneton moments. Irreducible representation and basis vectors used in the magnetic structure solution are listed. Basis vectors ($\psi$) are given in terms of the trigonal axes. Labeling follows the scheme used by Kovalev's tabulated works.[23] The magnetic moment for atom j is given by $m_j = C_1\psi_1 + C_2\psi_2$

Table 8: Irreducible representations, basis vectors, and magnetic structure model of $Pb_3TeCo_3P_2O_{14}$

| IR: | $\Gamma_1$ | k=(1/2, 0, -1/2) | | | $\Gamma_2$ | | |
|---|---|---|---|---|---|---|---|
| Atoms | $\Psi_1$ | $\Psi_2$ | $\Psi_3$ | $m_{total}$ $\mu_B$ | $\Psi_1$ | $\Psi_2$ | $\Psi_3$ |
| $Co_1$ | (2 0 0) | (0 0 2) | | 4.01 | (0 2 0) | | |
| $Co_2$ | (0 2 0) | | | 2.42 | (2 0 0) | (0 0 2) | |
| $Co_3$ | (2 0 0) | (0 0 2) | | 2.78 | (0 2 0) | | |
| $Co_4$ | (0 2 0) | | | 2.42 | (2 0 0) | (0 0 2) | |
| $Co_5$ | (2 0 0) | (0 0 2) | | 4.46 | (0 2 0) | | |
| $Co_6$ | (0 2 0) | | | 2.42 | (2 0 0) | (0 0 2) | |
| $Co_7$ | (1 0 0) | (0 1 0) | (0 0 1) | 3.76 | (1 0 0) | (0 1 0) | (0 0 1) |
| | (-1 0 0) | (0 1 0) | (0 0 -1) | | (1 0 0) | (0 -1 0) | (0 0 1) |
| $Co_8$ | (1 0 0) | (0 1 0) | (0 0 1) | 2.42 | (1 0 0) | (0 1 0) | (0 0 1) |
| | (-1 0 0) | (0 1 0) | (0 0 -1) | | (1 0 0) | (0 -1 0) | (0 0 1) |
| $Co_9$ | (1 0 0) | (0 1 0) | (0 0 1) | 4.90 | (1 0 0) | (0 1 0) | (0 0 1) |
| | (-1 0 0) | (0 1 0) | (0 0 -1) | | (1 0 0) | (0 -1 0) | (0 0 1) |
| $Co_{10}$ | (1 0 0) | (0 1 0) | (0 0 1) | 2.42 | (1 0 0) | (0 1 0) | (0 0 1) |
| | (-1 0 0) | (0 1 0) | (0 0 -1) | | (1 0 0) | (0 -1 0) | (0 0 1) |
| $Co_{11}$ | (1 0 0) | (0 1 0) | (0 0 1) | 3.10 | (1 0 0) | (0 1 0) | (0 0 1) |
| | (-1 0 0) | (0 1 0) | (0 0 -1) | | (1 0 0) | (0 -1 0) | (0 0 1) |
| $Co_{12}$ | (1 0 0) | (0 1 0) | (0 0 1) | 2.42 | (1 0 0) | (0 1 0) | (0 0 1) |
| | (-1 0 0) | (0 1 0) | (0 0 -1) | | (1 0 0) | (0 -1 0) | (0 0 1) |

Possible irreducible representation and basis vectors are listed. Data modeled with $\Gamma_1$, and moments on the cobalt atoms range from 2.42 to 4.9 Bohr magnetons.

TOC – Image

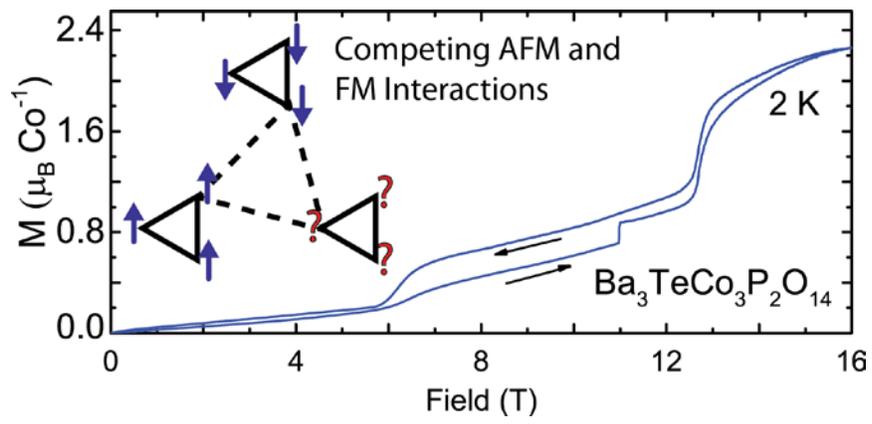